\documentclass[onecolumn, draft, 12pt]{IEEEtran}
\usepackage[final]{graphicx}
\usepackage{amsfonts}
\usepackage{amsmath}
\usepackage{relsize}
\usepackage{balance}
\usepackage{cite}
\usepackage{amssymb}
\usepackage[latin9]{inputenc}
\usepackage{centernot}
\usepackage{bm}
\usepackage{stmaryrd}
\usepackage{float}
\usepackage{array}
\usepackage[linesnumbered,ruled,vlined]{algorithm2e}
\usepackage[thmmarks]{ntheorem}
\usepackage{pdfpages}
\usepackage{bbold}
\usepackage{csquotes}
\usepackage{epstopdf}
\usepackage{url}
\usepackage{footnote}
\usepackage[bottom]{footmisc}
\usepackage{setspace}
\usepackage{calc}
\usepackage{mathptmx}
\DeclareMathAlphabet{\mathpzc}{OT1}{pzc}{m}{it}
\newlength{\depthofsumsign}
\newlength{\totalheightofsumsign}
\newlength{\heightanddepthofargument}
\SetKwComment{Comment}{$\triangleright$\ }{}
\newcommand{\nsum}[1][1.4]{
	\mathop{%
		\raisebox
		{-#1\depthofsumsign+1\depthofsumsign}
		{\scalebox
			{#1}
			{$\displaystyle\sum$}%
		}
	}
}

\newcommand\myeq{\mathrel{\overset{\makebox[0pt]{\mbox{\normalfont\tiny\sffamily def}}}{=}}}

\newtheorem{theorem}{Theorem}

\newtheorem{lemma}{Lemma}
\theoremheaderfont{\bfseries}
\theorembodyfont{\normalfont}
\theoremseparator{:}
\newtheorem{definition}{Definition}
\theoremsymbol{$\blacksquare$}

\renewcommand{\IEEEQED}{\IEEEQEDclosed}
\newcommand{\BigO}[1]{\ensuremath{\operatorname{O}\bigl(#1\bigr)}}
  
\newlength\myindent
\begin{document}
	\title{User-Base Station Association in HetSNets: Complexity and Efficient Algorithms}
	\author{Zoubeir~Mlika,~\IEEEmembership{Student~Member,~IEEE,}
		Mathew~Goonewardena,~\IEEEmembership{Student~Member,~IEEE,}
		Wessam~Ajib,~\IEEEmembership{Member,~IEEE,}
		and~Halima~Elbiaze,~\IEEEmembership{Member,~IEEE}
	}
	\maketitle
	\begin{abstract}
		This work considers the problem of user association to small-cell base stations (SBSs) in a heterogeneous and small-cell network (HetSNet). Two optimization problems are investigated, which are maximizing the set of associated users to the SBSs (the unweighted problem) and maximizing the set of weighted associated users to the SBSs (the weighted problem), under signal-to-interference-plus-noise ratio (SINR) constraints. Both problems are formulated as linear integer programs. The weighted problem is known to be NP-hard and, in this paper, the unweighted problem is proved to be NP-hard as well. Therefore, this paper develops two heuristic polynomial-time algorithms to solve both problems. The computational complexity of the proposed algorithms is evaluated and is shown to be far more efficient than the complexity of the optimal brute-force (BF) algorithm. Moreover, the paper benchmarks the performance of the proposed algorithms against the BF algorithm, the branch-and-bound (B\&B) algorithm and standard algorithms, through numerical simulations. The results demonstrate the close-to-optimal performance of the proposed algorithms. They also show that the weighted problem can be solved to provide solutions that are fair between users or to balance the load among SBSs.
	\end{abstract}
	
	\begin{IEEEkeywords}
		HetSNets, Heuristic algorithm, Brute-force, Branch-and-bound, NP-hard, Fairness, Load balancing.
	\end{IEEEkeywords}

	\IEEEpeerreviewmaketitle
	
	\section{Introduction\label{intro}}
	
	\subsection{Motivation and Research questions}
	
	In the last decade, mobile cellular networks have become popular among data users, which has led to a demand for increased capacity. In addition, cellular networks are becoming the main provider of voice and data services with high mobility even though the wireless local area networks (WLANs) can provide higher and less expensive data rates with relatively restricted mobility~\cite{interference_management}.
	In order for cellular networks to be more competitive with WLANs, resources must be adequately allocated to provide higher performance and better satisfy the requirements of users. To this end, small-cell base stations (SBSs) were introduced to form heterogeneous and small-cell network (HetSNet)~\cite{ekram}. In HetSNets, SBSs can provide operators an important increase in capacity at low capital expenditure~\cite{ekram}. They are low power, small range, and low price base stations~\cite{ekram,andrews_survey}. Despite the gains carried by SBSs, their deployment raises many challenges in HetSNets. In fact, a typical HetSNet is composed of a large number of SBSs that may exceed the number of users~\cite{5g}. Such densely deployment of SBSs has made the association of users to the SBSs, denoted as ``user-BS association'', a key challenge. Furthermore, HetSNets are interference-limited and hence the co-channel interference among SBSs and between SBSs and macro-cell BSs (MBSs) is a critical issue, which needs to be better managed to boost HetSNets capacity. It is clear that the user-BS association directly affects the interference and therefore can achieve enhanced capacity. Moreover, the basic user-BS association, which pairs the users to the SBSs that has the strongest signal, \textit{max-SINR}, has a low overall throughput because of a poor management of the interference~\cite{femto_past}. In this paper, we are interested in finding a user-BS association that increases the network capacity, defined by the number of associated user in one time-slot, such that the quality-of-service (QoS) of the associated users is guaranteed. Roughly speaking, we define this problem as follows: given a set of users (small-cell users (SUs) and a macro-cell user (MU)), a set of SBSs, one MBS, a QoS lower bound and channel gains between every pair of user-SBS, the question is to find a set of one-to-one association between the SUs and the SBSs with maximum cardinality such that the signal-to-interference-plus-noise ratio (SINR) of the SUs and of the MU are greater than the QoS lower bound.
	
	
	\subsection{Related works}
	
	Related work can be divided into: (\textit{i}) papers on distributed or centralized solutions of the user-BS association problem in HetSNets~\cite{Kuang, Qian, andrews_load, load_assoc,KaimingS,Gibbs}; and (\textit{ii}) papers on the link activation problem under SINR constraints~\cite{Olga, 5062048, Dinitz:2010:DAA:1833515.1833717} in wireless mesh networks. Next, we present the most recent related work on both directions.
	
	In~\cite{Kuang}, the authors study the resource allocation in HetSNets as a joint optimization problem of channel allocation, user-BS association, beam-forming and power control. It is solved using an iterative heuristic algorithm. Even though, the work shows that the relaxation of the combinatorial problem to a continuous one provides the optimal solution. The proof lacks of generality and it depends on the problem formulation. The proposed algorithm solves iteratively a convex approximation problems which leads to complex procedure. In~\cite{Qian}, the joint power allocation and user-BS association is modeled as a combinatorial optimization problem. The authors use Bender's decomposition to solve the modeled problem optimally and they propose heuristic algorithms. However, the proposed optimal method and the heuristic algorithms are highly complex. The user-BS association problem is solved in~\cite{andrews_load} and~\cite{load_assoc} jointly for fairness and load balancing. For instance, in~\cite{andrews_load}, the load of the BSs is balanced using a distributed algorithm based on the technique of dual decomposition. This work solves the user-BS association problem based on relaxation and rounding techniques which remove the combinatorial nature of the problem and render it easier to solve. Reference~\cite{KaimingS} solves the user-BS association in HetSNets based on a pricing scheme. The authors propose a dual coordinate descent method to solve the problem. The paper also extends the problem to the multiple-input-multiple-output (MIMO) case and optimizes the beam-forming variables. The optimization model is very similar to the one in~\cite{andrews_load}. The main difference with our paper is that the proposed solution is distributed with no optimal solution guarantee and there is no SINR constraints in the optimization problem. In~\cite{andrews_load,KaimingS}, multiple users have to be associated with one BS and all BSs have to be associated in the end which makes the optimization problem simpler. Reference~\cite{MingyiH} studies the joint problem of power control and user-BS association in HetSNets. The problem is modeled as a max-min fairness problem and the authors study its NP-hardness. First, the authors study the problem of maximizing the minimum SINR subject to power constraints and the association vector between the users and the SBSs. Second, they consider the additional constraints of one-to-one matching and of the SINRs guarantee. The first problem is shown to be NP-hard and the authors propose a two-stage fixed-point algorithm to solve it. The one-to-one matching problem is polynomial-time solvable and the authors propose an auction-based algorithm to solve it. Both problems are different from the one described in this paper. The main difficulty is to find the power and the association jointly. In the second problem, the critical assumption made is that the number of users is equal to the number of SBSs and all of them have to be associated (there is no maximization of the number of users) and therefore the authors reduce the problem to an assignment problem.
	
	On the other hand, the link activation problem consists of maximizing the size of the weighted set of activated links in one time-slot such that the per-link SINR constraint is satisfied~\cite{Olga}. The seminal work of Goussevskaia \textit{et al.}~\cite{Olga} shows that this problem,  \textit{one-slot scheduling}, is NP-hard. Note that there is some similarity between~\cite{Olga} and this work. In fact, our problem is somehow equivalent to the one-slot scheduling where the links are not established yet and they have no weights. Anyhow, the NP-hardness of the unweighted \textit{one-slot scheduling} is not investigated. Moreover, it is important to note that the NP-hardness of the weighted problem does not imply anything about the NP-hardness of the unweighted one~\cite{Crescenzi200110}. In~\cite{5062048, Dinitz:2010:DAA:1833515.1833717}, the authors provide approximation algorithms and game theoretic distributed solutions in order to solve the joint problem of unweighted \textit{one-slot scheduling} and power allocation under geometric SINR constraints. (There is no fading or any stochastic effects in the SINR.) 
	
	To the best of our knowledge, there is no NP-hardness studies of the user-BS association problem in the case of fixed transmit power where the objective is to maximize the set of associated users subject to the SINR constraints. Previous work have focused on simplified assumptions using relaxation and rounding techniques which remove the combinatorial nature of the user-BS association problem. Also, the interference constraints are often greatly simplified using graph-based models instead of SINR constraints. Moreover, large number of papers do not study the fairness and/or load balancing of the user-BS association problem in HetSNets which is an important aspect in wireless communications. Consequently, in this paper, we study the user-BS association problem under SINR constraints in HetSNets and we prove that it is an NP-hard problem and we study the fairness and load balancing of such problem. Finally, we develop efficient heuristic algorithms to solve it.
	
	The system performance metrics are throughput and fairness (or load balancing). Throughput is defined as the number of users that are successfully associated to the SBSs under SINR constraints and fairness is measured by the number of times a user is associated to the SBSs.
	
	\subsection{Contributions}
	This paper investigates two problems of unweighted and weighted user-BS association in an open access HetSNet. The unweighted problem maximizes the set of associated SUs to the SBSs whereas, the weighted problem maximizes the set of weighted associated SUs to the SBSs, under the SINR constraints.
	
	The main contributions of this paper are as follow:
	\begin{enumerate}
		
		\item We prove that the unweighted user-BS association problem is NP-hard.
		
		\item We develop efficient and simple heuristic algorithms to solve both unweighted and weighted user-BS association problems.
		
		\item We compare the developed algorithms against the brute-force (BF) optimal algorithm, the branch-and-bound (B\&B) algorithm, a standard user-BS association algorithm called max-SINR~\cite{femto_past,KaimingS} and a benchmark algorithm recently proposed in~\cite{MingyiH}.
		
		\item We evaluate the complexity of the proposed algorithms and the complexity of the BF algorithm. The complexity of the proposed heuristic algorithms is shown to be polynomial in time and hence is practical to implement in contrast to the exponential-time BF algorithm.
		
	\end{enumerate}
	
	\subsection{Organization}
	
	The rest of the paper is organized as follows. Section~\ref{sys} discusses the system model. The problem is formulated in Section~\ref{prob}. Section~\ref{hard} provides the proof of the NP-hardness of the unweighted user-BS association problem. Next, Section~\ref{opt} presents the BF and the B\&B optimal solutions. Section~\ref{sol} presents heuristic algorithms for the user-BS association for both unweighted and weighted problems, and analyzes the complexity of the algorithms. Section~\ref{simu} provides simulation results to compare the algorithms and Section~\ref{cl} concludes the paper.
	
	\section{System Model\label{sys}}
	
\begin{figure}[!ht]
		\centering
		\includegraphics[scale=1]{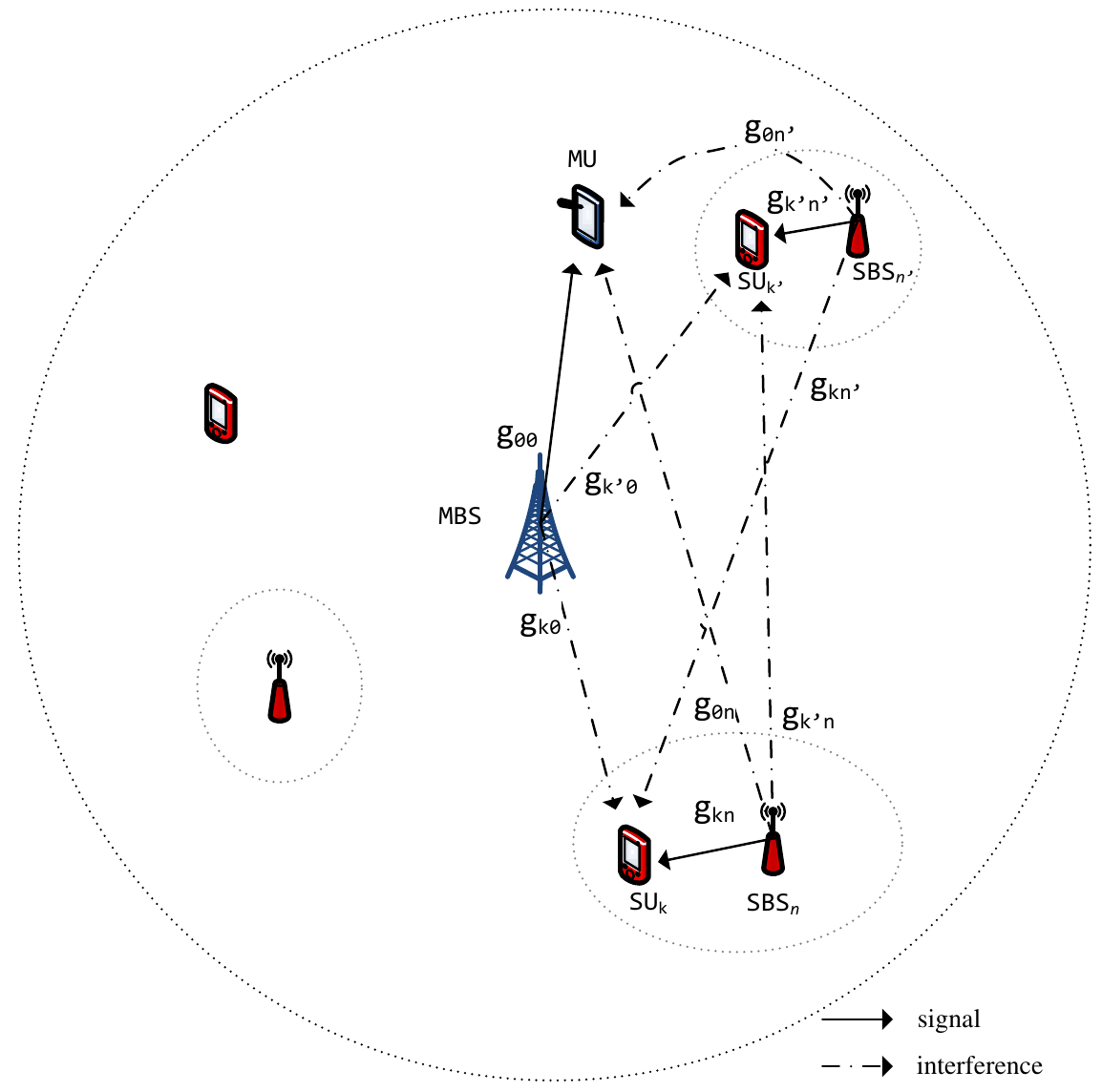}
		\centering	
		\caption{System Model}
		\label{sysmod}
	\end{figure}

	This paper considers the down-link transmission where all BSs transmit over a common frequency band. The network comprises a MBS, a macro-cell user (MU), several SBSs, and several small-cell users (SUs). We denote by $\mathcal{K}\myeq\{1, \cdots, K\}$ the set of SUs and by $\mathcal{N}\myeq\{1, \cdots, N\}$ the set of SBSs. For brevity, a SU and SBS are denoted simply by $k$ and $n$, respectively. The MBS and the MU are denoted by the index $0$. The MBS is located in the center of the cell which is modeled as a circle of radius $R$. SBSs, MU, and SUs are randomly located in this circle following independent two dimensional uniform distributions. An example of the system model is given in Fig.~\ref{sysmod}. 
	
	The wireless channel model includes path loss and Rayleigh fading. The channel between $k$ and $n$ is represented by $g_{kn}'\myeq h_{kn}\sqrt{(\frac{d_0}{d_{kn}})^\alpha}$, where $\alpha$ is the path loss coefficient, $d_{kn}$ is the distance between $k$ and $n$, $d_0$ is a reference distance at which the reference path loss is calculated (the reader can find more details in~\cite{Rappaport:2001:WCP:559977}), and $h_{kn}$ is the small-scale fading modeled as a zero-mean, complex Gaussian random variable with unit variance. Throughout the rest of the paper, we denote the channel gain by $g_{kn}\myeq|g_{kn}'|^2$. In this paper, each transceiver is equipped with a single antenna and one SU can be associated with one and only one SBS.
	
	For the mathematical formulation of the user-BS association problem, the binary variable $x_{kn}$ is defined as follows, for all $k\in\mathcal{K}$ and for all $n\in\mathcal{N}$:
	
	\begin{align*}
		x_{kn}\myeq
		\begin{cases}
			1 & \text{if $k$ is associated to $n$}\\ 
			0 & \text{otherwise.}
		\end{cases}
	\end{align*}
	
	The association vector $\mathbf{x}$, which represents the user-BS association solution, is defined as $\mathbf{x}\myeq\bigl[x_{11}, \ldots, x_{1N}\cdots x_{K1},\ldots, x_{KN}\bigr]^\mathrm{T}\in\{0, 1\}^{K\cdot N}$. Note that one SU can be associated with at most one SBS and one SBS can be associated with at most one SU and hence we have the following one-to-one association constraint on the vector $\mathbf{x}$: $\sum_{k\in\mathcal{K}}x_{kn}\leqslant1$ and $\sum_{n\in\mathcal{N}}x_{kn}\leqslant1$.
	
	The transmit power is normalized by the receiver noise power and the reference distance $d_0$. The SBSs transmit at constant SNR of $\gamma$. Although this assumption is a simplification to render the analysis tractable, it has been shown that constant transmit power method is useful in practice \cite{zvi}. The MBS transmits to its MU at constant SNR of $\gamma_0$. Then, the received SINR at $k$ from $n$ can be written as follows:
	
	\begin{align}\label{sinr:mn}
		\textrm{SINR}_{kn}\left(\mathbf{x}\right)\myeq\frac{\gamma g_{kn}x_{kn}}{1+\gamma_0g_{k0}+\nsum\limits_{\substack{k^\prime\in\mathcal{K}^\prime}}\nsum\limits_{\substack{n^\prime\in\mathcal{N}^\prime}}\gamma g_{kn^\prime}x_{k^\prime n^\prime}},
	\end{align}
	where $\mathcal{K}^\prime=\mathcal{K}\setminus\{k\}$ and $\mathcal{N}^\prime=\mathcal{N}\setminus\{n\}$.
	
	The SINR at the MU is given by:
	
	\begin{align}\label{sinr:0}
	\textrm{SINR}_{0}\left(\mathbf{x}\right)\myeq\frac{\gamma_0 g_{00}}{1+\nsum\limits_{\substack{k\in\mathcal{K}}}\nsum\limits_{\substack{n\in\mathcal{N}}}\gamma g_{0n}x_{kn}}.
	\end{align}
	
	The minimum required SINR threshold at any SU and at the MU are denoted by $\beta$ and $\beta_0$, respectively.  
	A user-BS association is feasible if and only if it meets the SINR threshold of the associated SUs and of the MU and if it satisfies the one-to-one association.
	
	\section{Problem Formulation \label{prob}}
	
	\subsection{Unweighted User-BS Association}
	
	This section formulates the unweighted user-BS association problem (the unweighted problem). The objective is to maximize the total number of associated SUs in the network subject to the constraints of the received SINR thresholds of the SUs and of the MU.
	
	The problem can be formulated as follows:
	
	\begin{subequations}
		\label{unw:pb}
		\begin{align}
		\underset{\mathbf{x}}{\text{maximize}}
		& \quad\nsum_{\substack{k\in\mathcal{K}}}\nsum_{\substack{n\in\mathcal{N}}}x_{kn}\label{eq:cost}\\
		\text{subject to} 
		& \quad\nsum_{k\in\mathcal{K}}x_{kn}\leqslant1,\;\forall\;n\in\mathcal{N},\label{cns:1}\\
		& \quad\nsum_{n\in\mathcal{N}}x_{kn}\leqslant1,\;\forall\;k\in\mathcal{K}, \label{cns:2}\\
		& \quad\textrm{SINR}_{kn}\left(\mathbf{x}\right)\geqslant\beta x_{kn},\;\forall\;k\in\mathcal{K},\;\forall\;n\in\mathcal{N},\label{cns:3}\\
		& \quad\textrm{SINR}_{0}\left(\mathbf{x}\right)\geqslant\beta_0,\label{cns:4}\\
		& \quad x_{kn}\in\left\{{0, 1}\right\},\;\forall\;k\in\mathcal{K},\;\forall\;n\in\mathcal{N}. \label{cns:5}
		\end{align}
	\end{subequations}
	
	Constraint \eqref{cns:1} ensures that a SBS associates to one SU whereas constraint \eqref{cns:2} ensures that a SU is associated with one SBS. Constraint \eqref{cns:3} guarantees that a SU associated with a SBS must have an SINR above the threshold $\beta$. To ensure the SINR threshold $\beta_0$ of the MU, constraint \eqref{cns:4} is introduced. Finally, constraint \eqref{cns:5} ensures that the association variable $x_{kn}$ is Boolean.
	
	Problem \eqref{unw:pb} can be written in matrix notation. Note that constraint \eqref{cns:3} is nonlinear due to the $\beta x_{kn}$ term on the right-hand side and the $x_{k^\prime n^\prime}$ in the denominator of the left-hand side. The $x_{kn}$ term dictates that the SINR threshold $\beta$ is met only if $k$ is associated to $n$. If it is not associated (i.e., $x_{kn}=0$), then the SINR threshold drops to zero and the constraint is satisfied with equality. Using the bigM technique~\cite{Schrijver:1986:TLI:17634}, constraint \eqref{cns:3} can be rewritten as below:
	
	\begin{align}
		\label{eq:M4}
		\frac{\gamma g_{kn}x_{kn}+M\left(1-x_{kn}\right)}{1+\gamma_0g_{k0}+\nsum\limits_{\substack{k^\prime\in\mathcal{K}^\prime}}\nsum\limits_{\substack{n^\prime\in\mathcal{N}^\prime}}\gamma g_{kn^\prime}x_{k^\prime n^\prime}}\geqslant\beta,
	\end{align}
	where $M$ is a sufficiently large number so that when $x_{kn}=0$, constraint \eqref{cns:3} is not violated and on the other hand if $x_{kn}=1$, the term $M\left(1-x_{kn}\right)$ is zero and therefore has no effect. This technique is well known in linear programming. It adds \enquote{artificial} variables to the original problem in order to find a feasible solution~\cite{Schrijver:1986:TLI:17634}. 
	
	The value of $M$ must satisfy the following for all $k\in\mathcal{K}$ and for all $n\in\mathcal{N}$:
	
	\begin{align}\label{mm:1}
	M \geqslant\beta+\beta\gamma_0g_{k0}+\nsum\limits_{\substack{k^\prime\in\mathcal{K}^\prime}}\nsum\limits_{\substack{n^\prime\in\mathcal{N}^\prime}}\beta\gamma g_{kn^\prime}x_{k^\prime n^\prime}.
	\end{align}
	
	Note that $M$ depends on $k$, $n$, and $\mathbf{x}$. Without loss of generality, we take the highest value of $M$ denoted by $M^*$:
	
	\begin{align}\label{mm:2}
	M^*\myeq\max_{k, n, \mathbf{x}}\left(\beta+\beta\gamma_0g_{k0}+\nsum\limits_{\substack{k^\prime\in\mathcal{K}^\prime}}\nsum\limits_{\substack{n^\prime\in\mathcal{N}^\prime}}\beta\gamma g_{kn^\prime}x_{k^\prime n^\prime}\right),
	\end{align}
	
	Hence, there exists $k^*\in\mathcal{K}$ and there exists $n^*\in\mathcal{N}$ such that equation \eqref{mm:2} is satisfied. Then, 
	
	\begin{equation}
	M^* \myeq \beta+\beta\gamma_0g_{k^*0}+\left(K-1\right)\beta\gamma\nsum\limits_{n^\prime\in\mathcal{N}^*}g_{k^*n^\prime},
	\end{equation}
	where $\mathcal{N}^*=\mathcal{N}\setminus\{n^*\}$.
	
	Using the previous value of $M^*$ and rearranging the terms, equation \eqref{eq:M4} is equivalent to:
	
	\begin{equation*}
	\left(\gamma g_{kn}-M^*\right)x_{kn}+M^*\geqslant\beta+\gamma_0g_{k0}\beta+\nsum\limits_{\substack{k^\prime\in\mathcal{K}^\prime}}\nsum\limits_{\substack{n^\prime\in\mathcal{N}^\prime}}\gamma\beta g_{kn^\prime}x_{k^\prime n^\prime}.
	\end{equation*}
	\[\Leftrightarrow\]
	\begin{align}\label{matrix2}
	\frac{M^*-\gamma g_{kn}}{M^*-\beta-\beta\gamma_0g_{k0}}x_{kn}+\nsum\limits_{\substack{k^\prime\in\mathcal{K}^\prime}}\nsum\limits_{\substack{n^\prime\in\mathcal{N}^\prime}}\frac{\gamma\beta g_{kn^\prime}}{M^*-\beta-\beta\gamma_0g_{k0}}x_{k^\prime n^\prime}\leqslant1.
	\end{align}
	
	Also, constraint \eqref{cns:4} can be rewritten as follows:
	
	\begin{align}
	\label{matrix4}
	\frac{\gamma_0 g_{00}}{\nsum\limits_{\substack{k\in\mathcal{K}}}\nsum\limits_{\substack{n\in\mathcal{N}}}\gamma g_{0n}x_{kn}+1}\geqslant\beta_0
	\Leftrightarrow
	\nsum\limits_{\substack{k\in\mathcal{K}}}\nsum\limits_{\substack{n\in\mathcal{N}}}\frac{\gamma g_{0n}\beta_0}{\gamma_0 g_{00}-\beta_0}x_{kn}\leqslant1.
	\end{align}
	
	With the above modifications, the unweighted user-BS association problem can be rewritten, in matrix form as follows:
	
	\begin{subequations}
		\label{unw:pbmatrixform}
		\begin{align}
		\underset{\mathbf{x}}{\text{maximize}}
		& \quad\mathbf{1}^\mathrm{T}\mathbf{x}\label{eq:costm}\\
		\text{subject to} 
		& \quad\mathbf{A}\mathbf{x}\leqslant\mathbf{1}, \label{matrixA}\\
		& \quad\mathbf{x}\in\left\{{0, 1}\right\}^q. \label{xbin}
		\end{align}
	\end{subequations}
	where $\mathbf{1}$ is the unitary vector of size $1\times q$ and $\mathbf{A}\in\mathbb{R}^{p\times q}$ is the matrix of sizes $p=K+N+K\cdot N+1$ and $q=K\cdot N$. The matrix $\mathbf{A}$ is defined by $\mathbf{A}=\left[a_{ij}\right]$ where $a_{ij}$ can be calculated from \eqref{cns:1}, \eqref{cns:2}, \eqref{matrix2}, and \eqref{matrix4}.
	
	\subsection{Weighted User-BS Association}
	
	This section introduces the more general problem of weighted user-BS association (the weighted problem) where each $k$ or $n$ in the network is prioritized by a weight. The problem is to maximize the number of weighted associated SUs subject to the constraints of the received SINR thresholds of the SUs and of the MU. The weights can add a degree of fairness to the users or balance the traffic load between the SBSs.
	
	The weighted problem can be formulated by defining a weight vector $\mathbf{w}$ which will be explained mathematically in the sequel. Thus, the weighted user-BS association problem is given below:
	
	\begin{subequations}
		\label{w:pbmatrixform}
		\begin{align}
		\underset{\mathbf{x}}{\text{maximize}}
		& \quad\mathbf{w}^\mathrm{T}\mathbf{x}\label{eq:costmw}\\
		\text{subject to} 
		& \quad\mathbf{A}\mathbf{x}\leqslant\mathbf{1}, \label{matrixAw}\\
		& \quad\mathbf{x}\in\left\{{0, 1}\right\}^q. \label{xbinw}
		\end{align}
	\end{subequations}
	
	Problem \eqref{w:pbmatrixform} is NP-hard~\cite{Olga}. The objective function of this problem is a linear combination of the variable $\mathbf{x}$ and a weight vector $\mathbf{w}$. When the vector $\mathbf{w}$ is set to one, the unweighted problem is obtained as in \eqref{unw:pbmatrixform}.
	
	\subsubsection*{Weights Design}
	The weights can be designed based on fair rate or fair time allocation \cite{5567009}. Since the problem involves the association of SUs to SBSs, we choose the fair time allocation. First, this paper considers the fairness between SUs and second, the weights are designed in order to provide fairness between SBSs. The fair time allocation between SBSs is also an important aspect and can be seen as a load balancing algorithm.
	
	Every $k$ (resp. $n$) is associated with a weight $w_{k}(t)$ (resp. $w_n(t)$) at time-slot $t$ which is, by definition, the reciprocal of number of times $k$ (resp. $n$) is associated during the previous period of $T$ time-slots, where $T$ is called the window size. Without loss of generality, we assume that the instant time $t$ is at least $T$, i.e., $t\geqslant T$. In other words, the weights are initialized for $t\leqslant T$. To ensure fairness between the SUs, the weights are calculated for every user based on the number of associations that occurred during the last $T$ time-slots and are given as follow for all $k\in\mathcal{K}$:
	
	\begin{align}
		\label{we2}
		w_{k}(t)\myeq\frac{1}{1+\nsum\limits_{n\in\mathcal{N}}\nsum\limits_{\tau=t-T+1}^tx_{kn}(\tau)},
	\end{align}
	where $x_{kn}(\tau)=1$, if $k$ is associated to $n$ at time $\tau$ and $x_{kn}(\tau)=0$ otherwise. For simplification, we omit the variable ($t$) from the weights when there is no possible confusion.
	The vector $\mathbf{w}$ denotes the weights vector and is given by $\\\mathbf{w}=\bigl[\underbrace{w_1, \ldots, w_1}_{\text{$N$~elements}} \cdots\allowbreak\underbrace{w_K, \ldots, w_K}_{\text{$N$~elements}}\bigr]^\mathrm{T}$. Similar procedure is followed in order to calculate the weights to ensure fairness between the SBSs at time $t$. Hence, for all $n\in\mathcal{N}$:
	
	\begin{align}
		\label{we3}
		w_{n}(t)\myeq\frac{1}{1+\nsum\limits_{k\in\mathcal{K}}\nsum\limits_{\tau=t-T+1}^tx_{kn}(\tau)},
	\end{align}
	and the corresponding weights vector $\mathbf{w}$ is given by $\mathbf{w}= \bigl[\underbrace{w_1, \ldots, w_N}_{\text{user~$1$}}\cdots\underbrace{w_1, \ldots, w_N}_{\text{user~$K$}}\bigr]^\mathrm{T}$.
	
	\section{NP-hardness\label{hard}}
	
	This section proves the NP-hardness of the unweighted user-BS association problem \eqref{unw:pbmatrixform}. The proof involves reducing a known NP-complete problem to problem \eqref{unw:pbmatrixform} in polynomial time. In this paper, the \textsc{Max Ones} problem \cite{doi:10.1137/S0097539799349948} is reduced to problem \eqref{unw:pbmatrixform}. The NP-hardness proof is divided into two parts. First, Lemma~\ref{lm} proves the NP-hardness of a sub-problem of \textsc{Max Ones} called \textsc{0-Valid Max Ones} using a reduction from the well known NP-complete problem \textsc{Set Cover}. Second, Theorem~\ref{thm} reduces \textsc{0-Valid Max Ones} to problem \eqref{unw:pbmatrixform}. 
	
	Please note that the NP-hardness of the weighted user-BS association does not apply anything about the NP-hardness of the unweighted user-BS association~\cite{Crescenzi200110}. Moreover, the structure of the matrix $\mathbf{A}$ cannot make the unweighted problem \eqref{unw:pbmatrixform} easy to solve because the matrix $\mathbf{A}$ is real-valued matrix and is not likely to be uni-modular.
	
	The symbols $\bigwedge$ (or $\land$), $\bigvee$ (or $\vee$), and $\neg$  denote the logical operators: disjunction, conjunction, and negation, respectively. The notation $P_1 \propto P_2$ is used to denote that problem $P_1$ is reducible in polynomial time to problem $P_2$.
	
	\begin{definition}[A binary constraint \cite{doi:10.1137/S0097539799349948}]\ \\
		A binary constraint is a function $f:\{0, 1\}^k\rightarrow\{0, 1\}$ for some $k \in \mathbb{N
		}$. We say that a binary constraint $f$ is satisfied by an input $\mathbf{s}\in\{0, 1\}^k$ if 
		$f(\mathbf{s})=1$.
	\end{definition}
	
	\begin{definition}[A \textit{0-valid} binary constraint \cite{doi:10.1137/S0097539799349948}]\ \\
		A binary constraint $f$ is \textit{0-valid} if $\mathbf{s}=\mathbf{0}$ and $f(\mathbf{s})=1$.
	\end{definition}
	
	\begin{definition}[\textsc{0-Valid Max Ones} problem \cite{doi:10.1137/S0097539799349948}]\label{def3}\ \\
		\textsc{Instance}: \textit{A \textit{0-valid} binary constraint $f(x_1, \cdots, x_n)$ of $n$ Boolean variables $x_1, x_2, \ldots, x_n$}.\\
		\textsc{Objective}: \textit{Decide if there are assignments to $x_1, x_2, \ldots, x_n$ that satisfy $f(\cdot)$ and find the one which has the most number of true variables, that to say $\max\left\{\sum_i x_i\right\}$}.
	\end{definition}
	
	\begin{definition}[\textsc{Set Cover} problem, NP-complete \cite{karp}]\label{scp}\ \\
		\textsc{Instance}: \textit{A set of $m$ elements called the universe. A finite family $\mathcal{J}$ of finite sets $S_j$ where $\mathcal{J}=\{\{S_j\}\,\forall\,j\}$, and a positive integer $k$}.\\
		\textsc{Objective}: \textit{Decide if there is a subfamily $\{T_h\} \subseteq \mathcal{J}$ that contains $e\leqslant k$ sets such that $\bigcup_h T_h =\mathcal{U}$}.
	\end{definition}
	
	Without loss of generality, an instance of \textsc{0-Valid Max Ones} problem is given by:
	
	\begin{align}\label{0valid}
	\underbrace{\left(\vee_{i\in\mathcal{S}_1}\neg x_i\right)}_{\text{clause 1}}\wedge\underbrace{\left(\vee_{i \in \mathcal{S}_2}\neg x_i\right)}_{\text{clause 2}}\wedge\cdots\wedge\underbrace{\vee_{i \in \mathcal{S}_L}\neg x_i}_{\text{clause L}}=\bigwedge_{l \in L}\bigvee_{i \in \mathcal{S}_l}\neg x_i,
	\end{align}
	where $\mathcal{S}_l$ for all $l\in L$, is a subset of $\{1, 2, \dotsc, n\}$. Equation \eqref{0valid} is the conjunction of disjunctions of $L$ clauses on the negated variables $\neg x_1, \cdots, \neg x_{|\mathcal{S}_l|}$.
	
	\begin{lemma}
		\label{lm}
		The \textsc{0-Valid Max Ones} problem is NP-hard.
	\end{lemma}
	
	\begin{IEEEproof}
		See Appendix~\ref{appendixA}.
	\end{IEEEproof}
	
	\begin{theorem}
		\label{thm}
		The unweighted user-BS association problem \eqref{unw:pbmatrixform} is NP-hard.
	\end{theorem}
	
	\begin{IEEEproof}
		See Appendix~\ref{appendixB}.
	\end{IEEEproof}
	
	The proof of Theorem~\ref{thm} is useful in wireless networks. In such networks, the user-BS association problem \eqref{unw:pbmatrixform} is often encountered. Unfortunately, due to Theorem~\ref{thm}, solving this problem optimally requires a BF method and needs vast computational capabilities unless $\mathrm{P=NP}$.
	The motivation behind the proof of Theorem~\ref{thm} is to find good algorithms that are less complex and perform close to the optimal solution.
	
	The next two sections present the proposed algorithms for the weighted and unweighted user-BS association problems along with the optimal solutions.
	
	\section{Optimal solutions \label{opt}}
	
	This section derives the optimal solutions for problems \eqref{unw:pbmatrixform} and \eqref{w:pbmatrixform}. The optimal solution can be calculated by two approaches, namely, the BF algorithm and the B\&B algorithm. The BF algorithm is based on the enumeration of all possible associations and picking the one with the best value. On the other hand, the B\&B algorithm reduces the search space, and hence the complexity, compared to the BF algorithm using the branching and the bounding approaches. These techniques are used as a reference for comparison against proposed algorithms. 
	
	In what follows, the complexity of the BF algorithm is derived for the unweighted user-BS association problem (denoted UBF) and for the weighted user-BS association problem (denoted WBF).
	
	\subsection{Unweighted User-BS Association}
	
	The basic steps of the UBF algorithm are the generation of all possible associations which are given by the enumeration of all combinations given by $C(K, N)$:
	\begin{align}
	C(K, N) = \nsum_{n=1}^{X_{(1)}}n!{X_{(1)}\choose n}{X_{(2)}\choose n},
	\end{align}
	where $\text{.}\choose\text{.}$ denotes the binomial coefficient and $X_{(1)}=\min\left(K, N\right)$ and $X_{(2)}=\max\left(K, N\right)$.
	
	Without loss of generality, let $N < K$, then: 
	
	\begin{align}
	\begin{split}
	C(K, N)\leqslant\nsum_{n=1}^N N!{K\choose N}{N\choose N}. \\
	\end{split}
	\end{align}
	
	From \cite{comb}, an upper bound of the binomial coefficient is given by $\displaystyle\left(\frac{n}{k}\right)^k \leqslant {n \choose k}\leqslant\frac{n^k}{k!}.$ Therefore:
	
	\begin{displaymath}
	C(K, N) \leqslant\begin{cases}
	K\cdot N^K & \text{if }N > K\,, \\
	N\cdot K^N & \text{if }N < K\,.
	\end{cases}
	\end{displaymath}
	
	The complexity of the UBF algorithm is denoted by \texttt{UBF-C}. Besides the enumeration of all possible combinations, the UBF algorithm runs through all the constraints, which is a matrix multiplication and has a complexity of $\BigO{p\cdot q}$, equivalently $\BigO{K^2\cdot N^2}$. Therefore, $\texttt{UOPT-C}\in\BigO{K^2\cdot N^2\cdot C(K, N)}\in\mathcal{X}_1$ where $\mathcal{X}_1$ is given by:
	
	\begin{align}
	\mathcal{X}_1\myeq
		\begin{cases}
			\BigO{K^3\cdot N^{K+2}} & \text{if }N > K\,, \\
			\BigO{N^3\cdot K^{N+2}} & \text{if }N < K\,, \\
			\BigO{N^5\cdot N!}  & \text{if }N = K\,.
		\end{cases}
	\end{align}
	
	\subsection{Weighted User-BS Association}
	
	The WBF algorithm follows mainly the same principle as of the UBF algorithm with a slight difference. After the generation of all combinations, each step calculates the weights (for a fixed $t$) for those combinations that satisfy the constraints and picks the one with the maximum value. The constraints verification requires $\BigO{p\cdot q}$, equivalently $\BigO{K^2\cdot N^2}$ and the calculation of the weights of those solutions requires $\BigO{q}$, equivalently $\BigO{K\cdot N}$, which gives a complexity of $\BigO{K^3\cdot N^3}$. Therefore, the complexity of the WBF algorithm, denoted by \texttt{WBF-C}, is $\texttt{WBF-C}\in\BigO{K^3\cdot N^3\cdot C(K, N)}\in\mathcal{X}_2$ where $\mathcal{X}_2$ is given by:
	
	\begin{align}
		\mathcal{X}_2\myeq
		\begin{cases}
			\BigO{K^4\cdot N^{K+3}} & \text{if }N > K\,, \\
			\BigO{N^4\cdot K^{N+3}} & \text{if }N < K\,, \\
			\BigO{N^7\cdot N!}  & \text{if }N = K\,.
		\end{cases}
	\end{align}
	
	\subsection{Branch-and-Bound Solution}
	The B\&B algorithm is a well known method to solve discrete and combinatorial optimization problems \cite{Schrijver:1986:TLI:17634}. It enumerates all possible solutions in a rooted tree. Then, it explores the branches of the rooted tree and estimates an upper and lower bounds on the optimal solution.
	
	In this paper, the B\&B algorithm with the CPLEX solver~\cite{cplex} is used to calculate the optimal solutions of problems \eqref{unw:pbmatrixform} and \eqref{w:pbmatrixform}. The computational complexity of this algorithm is exponential in the worst case. We are unable to give an analytical expression of the complexity since it is not known how such an algorithm is implemented. However, this algorithm works well in practice as experiments show and as suggested in the documentation of the CPLEX solver~\cite{cplex}. Hence, only the complexity of BF algorithm is provided. Even though, the B\&B algorithm has an exponential complexity in the worst case, it works faster than the BF algorithm in practice. To have an idea about the difference between the computational complexity of the BF algorithm and the B\&B algorithm, let us see an illustrative example. If the input is fixed to $K=10$, $N=6$ and the matrix $\mathbf{A}$ is a priori known then, based on a computer characterized by \enquote{Intel(R) Core(TM) i7-3770 CPU @ 3.40 GHz 3.40 GHz}, the running time for the BF algorithm is approximately equals to $4$ seconds whereas it is approximately equals to $0.1$ seconds for the B\&B algorithm.
	
	\section{Heuristic solutions \label{sol}}
	
	This section describes the proposed algorithms to solve both problems \eqref{unw:pbmatrixform} and \eqref{w:pbmatrixform}, which consist of simple but efficient greedy algorithms.
	
	We define $\mathbf{G}=[g_{kn}]$ for all $k\in\mathcal{K}\cup\{0\}$ and for all $n\in\mathcal{N}\cup\{0\}$ to represents the matrix of channel gains. In the pseudo-codes of the algorithms, we adopt the following notation $x,y\gets z,t$ to assign $z$ to $x$ and $t$ to $y$.
	
	\subsection{Unweighted Maximum Relative Channel Gain (UMRCG)}
	
	\begin{algorithm}
		\DontPrintSemicolon
		\KwIn{Network parameters: $\mathbf{G}$, $K$, $N$, $\gamma$, $\beta$, $\gamma_0$, $\beta_0$}
		\KwOut{A near optimal solution: $\mathbf{a}$}
		Create the matrix $\mathpzc{U}$ according to \eqref{eq:l1}\;\label{algo:1}
		$\mathbf{a},\mathsf{p}\gets\left[\;\right],0$\;\label{algo:2}
		\While{$\mathsf{p}<K\cdot N$}
		{\label{algo:3}
			$(k,n)\gets$ \textit{min}$(\mathpzc{U})$\tcp*[f]{Get the indexes of the smallest element in $\mathpzc{U}$.}\;\label{algo:4}
			$\mathbf{a}[n]\gets k$\tcp*[f]{Associate $n$ to $k$.}\;\label{algo:5}
			$(\mathsf{sinr},\mathsf{sinr}_0)\gets$ \textit{SINR}$(\mathbf{a})$\tcp*[f]{Return the SINRs of SUs and of MU.}\;\label{algo:6}
			$\mathsf{bool}\gets$ \textup{false}\;\label{algo:7}
			\For{$j = 1$ \textup{\textbf{to}} \textup{length}$(\mathsf{sinr})$}
			{\label{algo:8}
				\If{$\mathsf{sinr}[j]\geqslant\beta$ \textup{\textbf{and}} $\mathsf{sinr}_0\geqslant\beta_0$}
				{\label{algo:9}
					$\mathsf{bool}\gets$ \textup{true}\;\label{algo:10}
				}
				\Else
				{\label{algo:11}
					$\mathbf{a}[n],\mathsf{bool}\gets [\;],$ \textup{false}\tcp*[f]{Dissociate $n$ from $k$.}\;\label{algo:12}
					\textbf{break}\;\label{algo:13}
				}
			}
			\If{$\mathsf{bool}$ \textup{\textbf{is} true}}
			{\label{algo:14}
				\textit{eliminate}$(k,n)$\tcp*[f]{Do not re-assign neither $n$ nor $k$ next.}\;\label{algo:15}
			}
			$\mathsf{p}\gets \mathsf{p}+1$\;\label{algo:16}
		}
		
		\Return{$\mathbf{a}$}\label{algo:17}
		\caption{{\sc UMRCG}}
		\label{algo:umrcg}
	\end{algorithm}
	The proposed algorithm to solve the unweighted problem is denoted by UMRCG and is given in Algorithm~\ref{algo:umrcg}. It solves the unweighted user-BS association problem heuristically based on a greedy method. \textbf{First, in line~\ref{algo:1}, it creates a matrix $\mathpzc{U}=[u_{kn}]$ for all $k$ and $n$ as follows:}
	\begin{align}
		\label{eq:l1}
		\mathpzc{u}_{kn}=\frac{g_{kn}}{\nsum\limits_{k^\prime\neq k}g_{k^\prime n}}.
	\end{align}
	Note that this matrix plays a key role in the proposed algorithm. In fact, $u_{kn}$ represents the receivable signal power of $k$ divided by the sum of receivable signal powers of other $k'\neq k$. Hence, $\mathpzc{u}_{kn}$ can be seen as the inverse of the price of associating $k$ to $n$.
	
	After the creation of the matrix $\mathpzc{U}$, Algorithm~\ref{algo:umrcg}, in line~\ref{algo:2}, initializes the association vector $\mathbf{a}$ to the empty vector and the counter $\mathsf{p}$ to zero. The association vector defines the choice of each $n\in\mathcal{N}$, i.e., $\mathbf{a}[n]=k$ means that $k$ is associated to $n$. Next, line~\ref{algo:3} traverses the whole matrix $\mathpzc{U}$ inside the while loop. At every iteration in this loop, Algorithm~\ref{algo:umrcg} in line~\ref{algo:4} finds, using the function \textit{min}$(\cdot)$, the indexes $k$ and $n$ of the smallest element of $\mathpzc{U}$. Then, the algorithm associates $k$ to $n$. According to the association vector $\mathbf{a}$ created so far, the algorithm calculates the SINRs, using the function \textit{SINR}$(\cdot)$, of the SUs and of the MU. For every calculated SINR, the algorithm tests whether it is greater or equal than the thresholds $\beta$ and $\beta_0$ as given in line~\ref{algo:8}. If the association vector does not violate any SINR constraint so far, a Boolean variable $\mathsf{bool}$ is assigned a true value. If not, $k$ is dissociated from $n$, $\mathsf{bool}$ is set to false and the loop is broken. In line~\ref{algo:14}, if $\mathsf{bool}$ is true, which means that the newly association $\mathbf{a}[n]=k$ is valid for all associated pairs of SU-SBS, then the corresponding $k$ and $n$ cannot be used for any further association in the subsequent iterations. Therefore, the function \textit{eliminate}$(\cdot,\cdot)$, in line~\ref{algo:15}, sets the row $k$ and the column $n$ of $\mathpzc{U}$ to a very large number to prevent choosing them next. Note that this guarantees that constraints~\eqref{cns:1} and~\eqref{cns:2} are not violated. In line~\ref{algo:16}, the counter $\mathsf{p}$ is updated and the while loop continues. Finally, when all the elements of the matrix $\mathpzc{U}$ are evaluated, the algorithm halts and returns a sub-optimal user-BS association vector $\mathbf{a}$.
	
	The UMRCG algorithm runs in polynomial time. The creation of the matrix $\mathpzc{U}$ requires $\BigO{K\cdot N}$ if we store the sum $s_n=\sum_{k=1}^{K}g_{kn}$ in a list of $N$ elements and we calculate $\mathpzc{u}_{kn}$ as $\mathpzc{u}_{kn}=\frac{g_{kn}}{s_n-g_{kn}}$ for all $k$ and $n$. The while loop requires $\BigO{K\cdot N}$ in the worst case. The function \textit{min}$(\cdot)$ requires $\BigO{K\cdot N}$. The \textit{SINR} function needs to calculate the SINR of every associated pairs SU-SBS and of the pair MU-MBS which requires $\BigO{K\cdot N}$ by the same technique used to create the matrix $\mathpzc{U}$. Line~\ref{algo:8} through line~\ref{algo:13} require $\BigO{N}$ in the worst case. At the end, the function \textit{eliminate}$(\cdot, \cdot)$ goes through the row $k$ and the column $n$ which requires $\BigO{K+N}$. Finally the overall complexity of the UMRCG algorithm, denoted by \texttt{UMRCG-C}, is given in the worst case by $\BigO{K\cdot N+2\cdot K^2\cdot N^2+K\cdot N^2+2\cdot K\cdot N^2},$ which can be simplified to:
	
	\begin{equation}
		\texttt{UMRCG-C}\in\BigO{K^2\cdot N^2}.
	\end{equation}
	
	\subsection{Weighted Maximum Relative Channel Gain (WMRCG)}
	
	\begin{algorithm}
		\DontPrintSemicolon
		\KwIn{Network parameters: $T$, $\mathbf{G}$, $K$, $N$, $\gamma$, $\beta$, $\gamma_0$, $\beta_0$}
		\KwOut{A near optimal solution $\mathbf{a}$}
		Calculate the weights in a period of size $T$ called the window \tcp*[f]{Return the weights of the SBSs and the SUs calculating during $T$.}\;\label{algow:1}
		\For{$t\geqslant T$}
		{\label{algow:2}
			Create the matrix $\mathpzc{W}$ according to~\eqref{eq:l2}\;\label{algow:3}
			$\mathbf{a},\mathsf{p}\gets\left[\;\right],0$\;\label{algow:4}
			\While{$\mathsf{p}<K\cdot N$}
			{\label{algow:5}
				$(k, n)\gets$ \textit{min}$(\mathpzc{W})$\;\label{algow:6}
				$\mathbf{a}[n]\gets k$\;\label{algow:7}
				$(\mathsf{sinr},\mathsf{sinr}_0)\gets$ \textit{SINR}$(\mathbf{a})$\;\label{algow:8}
				$\mathsf{bool}\gets$ \textup{false}\;\label{algow:9}
				\For{$j = 1$ \textup{\textbf{to}} \textup{length}$(\mathsf{sinr})$}
				{\label{algow:10}
					\If{$\mathsf{sinr}[j]\geqslant\beta$ \textup{\textbf{and}} $\mathsf{sinr}_0\geqslant\beta_0$}
					{\label{algow:11}
						$\mathsf{bool}\gets\text{true}$\;\label{algow:12}
					}
					\Else
					{\label{algow:13}
						$\mathbf{a}[n],\mathsf{bool}\gets[\;]$, \textup{false}\;\label{algow:14}
						\textbf{break}\;\label{algow:15}
					}
				}
				\If{$\mathsf{bool}$ \textup{\textbf{is} true}}
				{\label{algow:16}
					\textit{eliminate}$(k,n)$\;\label{algow:17}
				}
				$\mathsf{p}\gets\mathsf{p}+1$\;\label{algow:18}
			}
			Move the window $T$\;\label{algow:19}
			Update the weights according to~\eqref{we2} or~\eqref{we3}\;\label{algow:20}
		}
		\Return{$\mathbf{a}$}\label{algow:21}
		\caption{{\sc WMRCG}}
		\label{algo:wmrcg}
	\end{algorithm}
	
	The proposed algorithm to solve the weighted problem is denoted by WMRCG. It is divided into two steps. The first step, in line~\ref{algow:1}, is the calculation of the weights according to~\eqref{we2} or~\eqref{we3} during the window $T$. The second step, from line~\ref{algow:2} to line~\ref{algow:20}, the algorithm WMRCG solves the weighted user-BS association problem using a procedure similar to the one described in the UMRCG algorithm. The main differences between the UMRCG algorithm and the WMRCG algorithm are the criterion in line~\ref{algow:3} and the update of the weights in lines~\ref{algow:19} and~\ref{algow:20}. In line~\ref{algow:3}, the algorithm WMRCG creates the matrix $\mathpzc{W}=[\mathpzc{w}_{kn}]$ of SU-SBS pairs for all $k$ and $n$ as follows:
	\begin{align}
		\label{eq:l2}
		\mathpzc{w}_{kn}=\frac{w_{k|n}g_{kn}}{\nsum\limits_{k^\prime\neq k}g_{k^\prime n}},
	\end{align}
	\textbf{where $w_{k|n}$ is $w_k$ or $w_n$, depending on whether to balance the load among the SBSs or to be fair between the SUs as discussed previously.}
	
	Likewise, the WMRCG algorithm runs in polynomial time for a fixed period of time. On the one hand, the first step of calculating the weights needs to go through the association vector during the window of $T$ time-slots and calculates how many times $k$ (resp. $n$) has been associated according to~\eqref{we2} (resp.~\eqref{we3}). This requires $\BigO{K\cdot N\cdot T}$.
	On the other hand, similarly to the previous analysis of the UMRCG algorithm, the complexity of the second step of the WMRCG requires $\BigO{K^2\cdot N^2}$. Finally, the overall complexity of the WMRCG algorithm, denoted by \texttt{WMRCG-C}, is given by:
	\begin{equation}
		\texttt{WMRCG-C} \in \BigO{N\cdot K\cdot T + K^2\cdot N^2}.
	\end{equation}
	
	\begin{table}[!ht]
		\centering
		\begin{tabular}{|l|p{6cm}|p{6cm}|}                                                 \hline
			Algorithm        & Complexity, $K<N$                      & Example         \\ \hline
			\texttt{UBF-C}   & $\BigO{K^3\cdot N^{K+2}}$              & $4\cdot10^{18}$ \\ \hline
			\texttt{WBF-C}   & $\BigO{K^4\cdot N^{K+3}}$              & $8\cdot10^{20}$ \\ \hline
			\texttt{UMRCG-C} & $\BigO{N^2\cdot K^2}$                  & $4\cdot10^{4}$  \\ \hline
			\texttt{WMACG-C} & $\BigO{N\cdot K\cdot T + K^2\cdot N^2}$& $24\cdot10^{4}$ \\ \hline
		\end{tabular}
		\vspace{1mm}
		\caption{Complexity of the algorithms}
		\label{table1}
	\end{table}
	
	Table~\ref{table1} summaries the complexity of the proposed algorithms and of the optimal algorithms. We see that UMRCG and WMRCG have very low complexity compared to the UBF and WBF. In fact the complexity of both algorithms is quadratic in either $K$ or $N$. Notice that UMRCG and WMRCG have almost the same order of complexity unless $T$ is of the same order as $K^2$ and $N^2$. As an illustrative example, we set $K=10$, $N=20$, and $T=1000$ in the third column of table~\ref{table1}. We see the huge difference in the computational complexity between $4\cdot10^{18}$ of the UBF algorithm and $4\cdot10^{4}$ of the UMRCG algorithm. These results demonstrate the advantage of using heuristic algorithms and show how the proposed algorithms are computationally simple. 
	
	Since both the weighted and unweighted problems are NP-hard, there are no polynomial time algorithms that solve them optimally unless $\mathrm{P=NP}$. Therefore, our proposed algorithms can be used and implemented to solve such problems in real scenarios. Note, however, that the proposed algorithms does not approximate the optimal solution theoretically, i.e., we cannot argue that, for any instance of size $\ell$ of both problems, the ratio between the solutions of the proposed algorithms and the optimal algorithm is at least $\rho(\ell)\leqslant1$. A rigorous analysis of the performance ratio of the greedy algorithms against the optimal would be an extensive work that cannot be integrated with this work. In fact, one has to study the hardness of approximation of the user-BS association problem first in order to guarantee the existence of an approximation algorithm. In~\cite{olga:2}, the authors proved that it is NP-hard to approximate the one-slot scheduling problem under the abstract SINR constraints (which is very similar to the unweighted user-BS association problem) to within $n^{1-\varepsilon}$, for any $\varepsilon>0$. Therefore, the user-BS association problem is apparently hard to approximate. Such contribution is left for future work.
	
	\section{Simulation Results\label{simu}}
	
	In this section, the performance of the proposed algorithms is demonstrated by simulations. It is assumed that the path loss coefficient is $\alpha=4$ which is a typical value in cellular networks~\cite{Rappaport:2001:WCP:559977}, and the radius of the circle where the SBSs are located is $R = 20$ m~\cite{andrews_survey}. Unless otherwise specified, the transmit SNR of the MBS and of the SBSs are set to $\gamma_0 = 40$ dB and $\gamma = 20$ dB, respectively. The SINR thresholds used for the MU and for the SUs are given respectively by $\beta_0 = 0$ dB and $\beta = 1$ dB, and the number of SUs is $K = 10$. The WMRCG is executed with a window of size $T=50$. The B\&B algorithm is implemented using the OPTI Toolbox~\cite{CW12a} under MATLAB using the IBM ILOG CPLEX solver~\cite{cplex}.
	
	\begin{figure}[!h]
		\centering
		\includegraphics[scale=0.8]{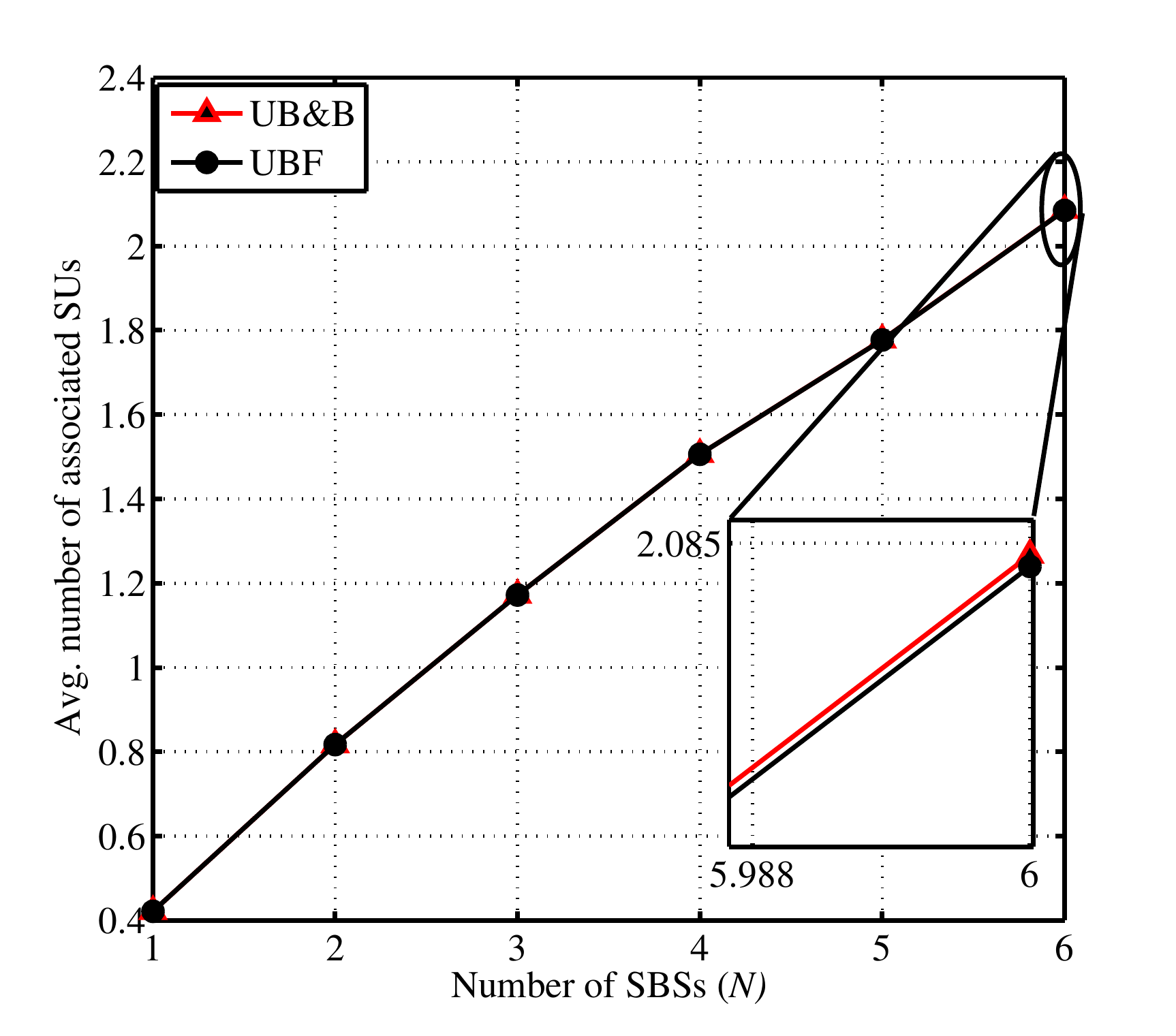}
		\caption{Performance of unweighted optimal solutions (UB\&B and UBF algorithms).}
		\label{opt_bf_opt_bnb}
	\end{figure}
	Fig. \ref{opt_bf_opt_bnb} compares UBF algorithm and B\&B algorithm for the unweighted problem (denoted UB\&B). We see that UBF slightly outperforms UB\&B especially when $N$ is high. When $N=6$, UBF solution is $.009\%$ far away from UB\&B one. However, this small difference is generally due to the floating points representation errors of B\&B algorithm. Fig.~\ref{opt_bf_opt_bnb} along with the complexity analysis in Table~\ref{table1} illustrate that UB\&B algorithm allows us to obtain tight-to-optimal performance with relatively low computational complexity. This motivates us to use B\&B algorithm in our next simulations.
	
	\begin{figure}[!h]
		\centering
		\includegraphics[scale=0.8]{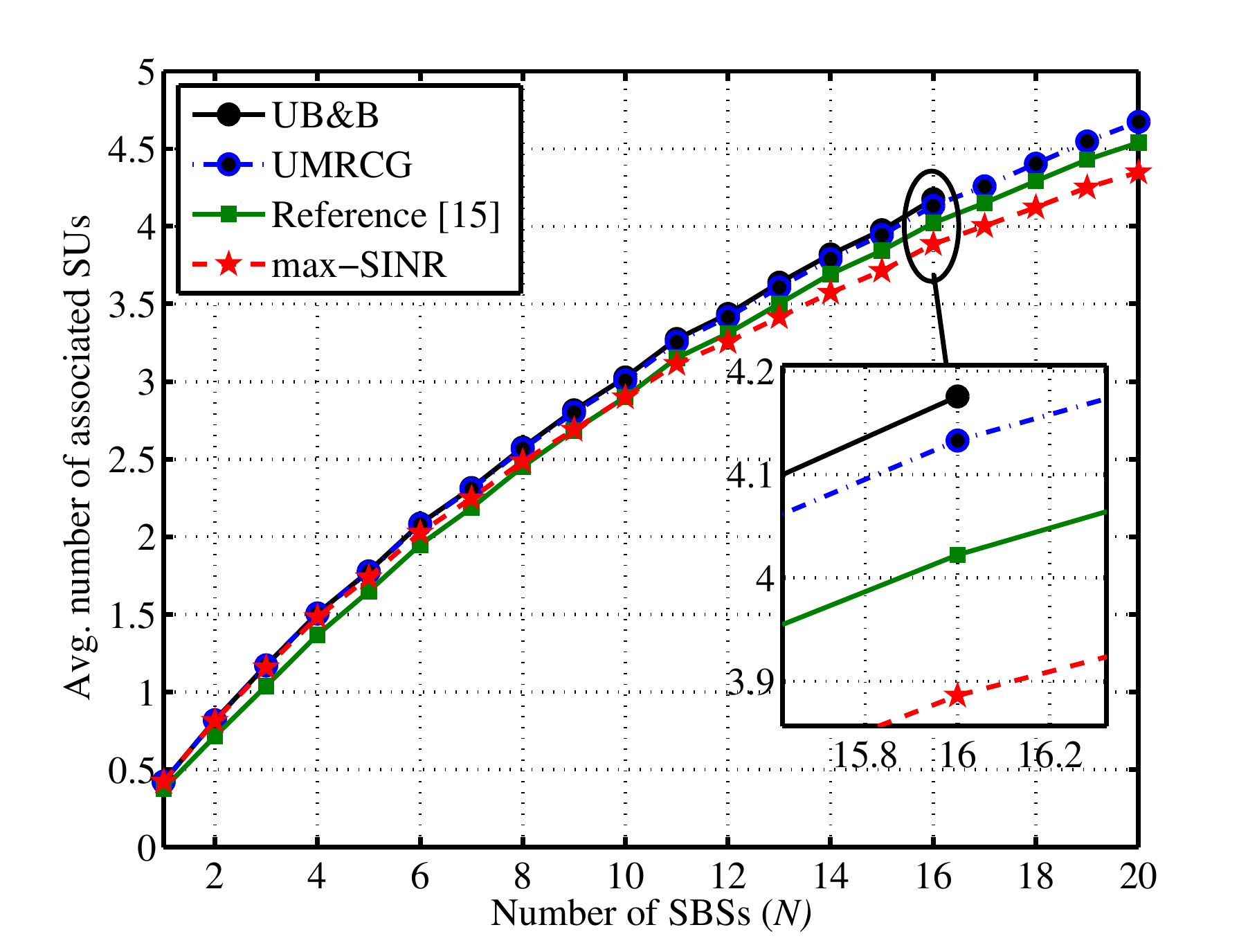}
		\caption{Performance of the UB\&B algorithm and the UMRCG algorithm with comparison to the max-SINR algorithm and reference~\cite{MingyiH} for the unweighted user-BS association problem.}
		\label{opt_grd_u}
	\end{figure}
	Fig. \ref{opt_grd_u} shows the average number of associated SUs for the unweighted user-BS association problem \eqref{unw:pbmatrixform}. It compares UB\&B, UMRCG, a benchmark algorithm denoted by max-SINR and a recently proposed algorithm~\cite{MingyiH}. (The algorithm in~\cite{MingyiH} is adapted to our situation.) In the max-SINR algorithm, each SU is associated to the strongest SBS signal it receives whereas in the criterion used in~\cite{MingyiH} each SU is associated to an SBS according to the sum of the received interference. This criterion works well for~\cite{MingyiH} since all the SBSs are associated in the end and therefore the sum of the received interference is not predicted correctly. We see that UMRCG algorithm has very close performance to the optimal solution. E.g., UMRCG solution is $.958\%$ far away from UB\&B solution when $N=16$. Furthermore, the proposed UMRCG algorithm outperforms max-SINR algorithm since the latter does not provide a good interference management among the BSs. Moreover, our proposed algorithm beats the algorithm in~\cite{MingyiH} since in our proposed algorithm, some of the SBSs may not be associated and therefore the amount of interference is overestimated. Note that, the performance of proposed algorithms depend on the number of SUs $K$ and SBSs $N$ (as shown in Fig.~\ref{opt_grd_u}), on the transmit SNR and on the SINR thresholds. Next, we demonstrate the effect of the transmit powers and the thresholds on the performance of the proposed solutions.
	
	\begin{figure}[!h]
		\centering
		\includegraphics[scale=0.8]{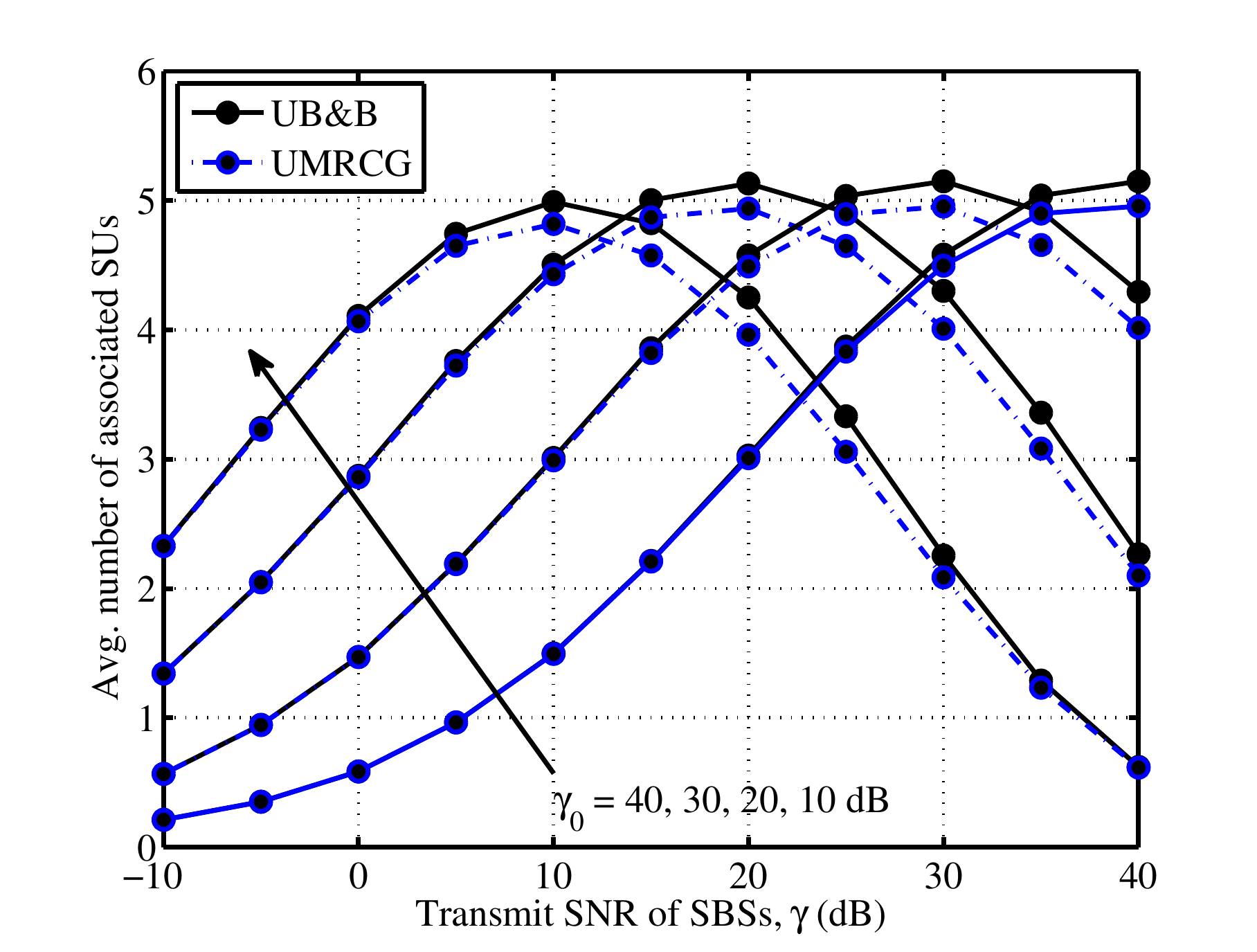}
		\caption{Performance of the the UB\&B algorithm and the UMRCG algorithm versus the transmit SNR of the SBSs $\gamma$ for different transmit SNR of the MBS $\gamma_0$. $N = 10$.}
		\label{gamma}
	\end{figure}
	Fig. \ref{gamma} plots the average number of associated SUs versus the transmit SNR $\gamma$ of the SBSs. When $\gamma$ grows, the SINR received at the SUs grows and more SUs are expected to be associated which is illustrated in the first part of the x-axis in Fig.~\ref{gamma} when the curves increase. When $\gamma$ continues to grow, the interference at the MU grows too and becomes harmful. Hence, the MU is not satisfied and the average number of SUs must decrease to respect the MU's QoS. This is illustrated in the second part of the x-axis in Fig. \ref{gamma} when the curves dip. Notice that for high $\gamma$, if $\gamma_0$ increases, then the average number of associated SUs increases. E.g., we observe that when $\gamma=40$ dB, the average number of associated SUs increases from $.6$ to approximately $5$ as $\gamma_0$ increases from $10$ dB to $40$ dB. On the other hand, for smaller $\gamma$, if $\gamma_0$ increases, then less SUs are associated. Therefore, for a given value of the transmit SNR of the MBS, $\gamma_0$, there is an optimum value of the transmit SNR of the SBSs, $\gamma$, to be used in order to maximize the number of associated SUs. Finally, we can see that the proposed algorithm UMRCG still gives close-to-optimal performance for different values of transmit SNR. Fig.~\ref{gamma} shows the worst case ratio between the UMRCG solution and the UB\&B one is at most $5\%$.
	
	\begin{figure}[h!]
		\centering
		\includegraphics[scale=0.8]{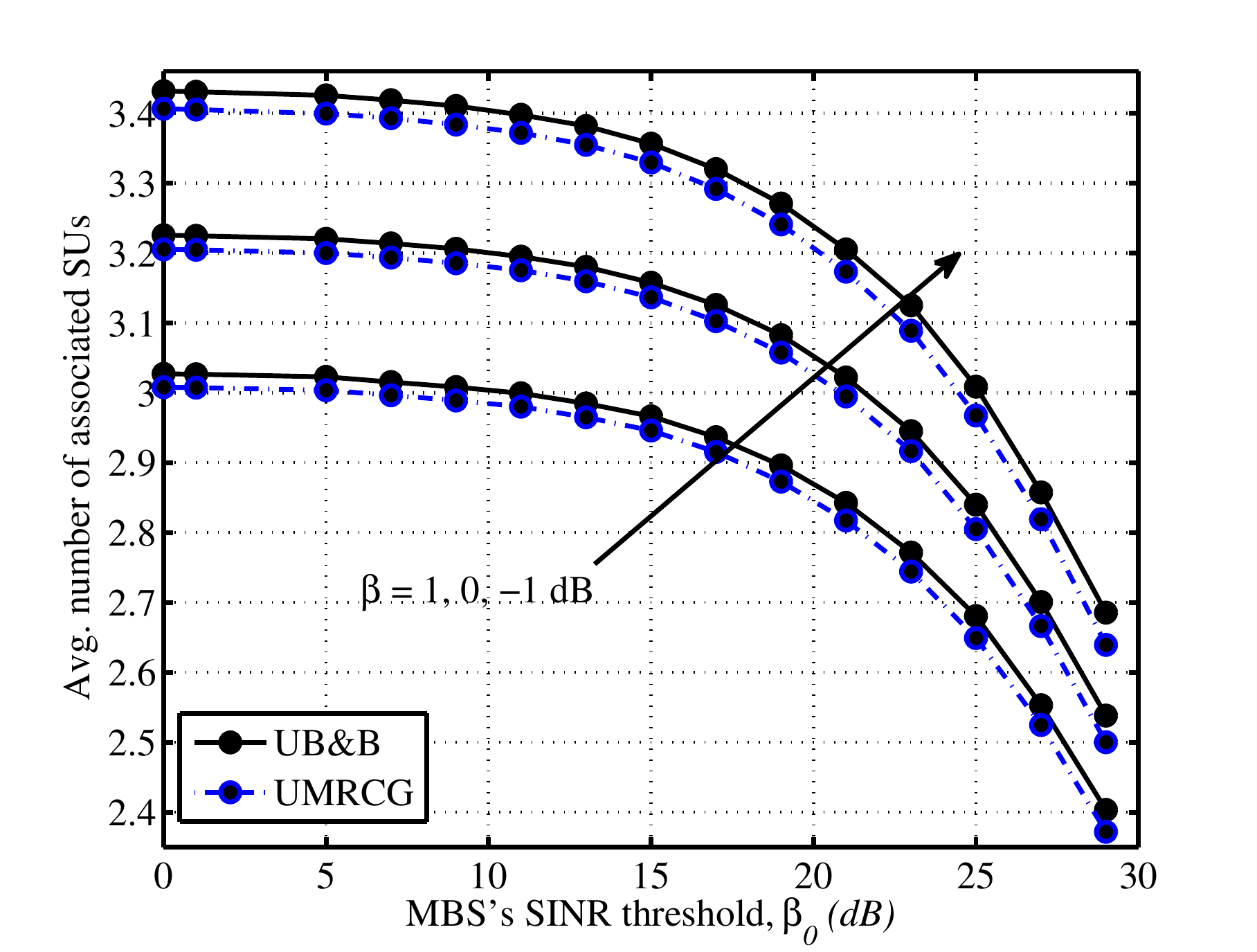}
		\caption{Performance of the UB\&B algorithm and the UMRCG algorithm versus the SINR thresholds of the MBS $\beta_0$ for different SINR threshold of the SBSs $\beta$. $N = 10$.}
		\label{Gamma_0}
	\end{figure}
	Fig. \ref{Gamma_0} depicts the effect of SINR thresholds of SUs and of the MU. The average number of associated SUs decreases when the thresholds increase. When $\beta_0$ gets smaller, the QoS of the MU is satisfied more often and hence more SUs get associated. Furthermore, when $\beta_0$ becomes higher, the number of associated SUs decreases dramatically regardless of the value of $\beta$. It is also important to notice that the ratio between UB\&B solution and UMRCG solution varies slightly as a function of $\beta_0$ and $\beta$. This ratio is still small though, which illustrates the accuracy of the proposed heuristic solution.
	
	\begin{figure}[h!]
		\centering
		\includegraphics[scale=0.8]{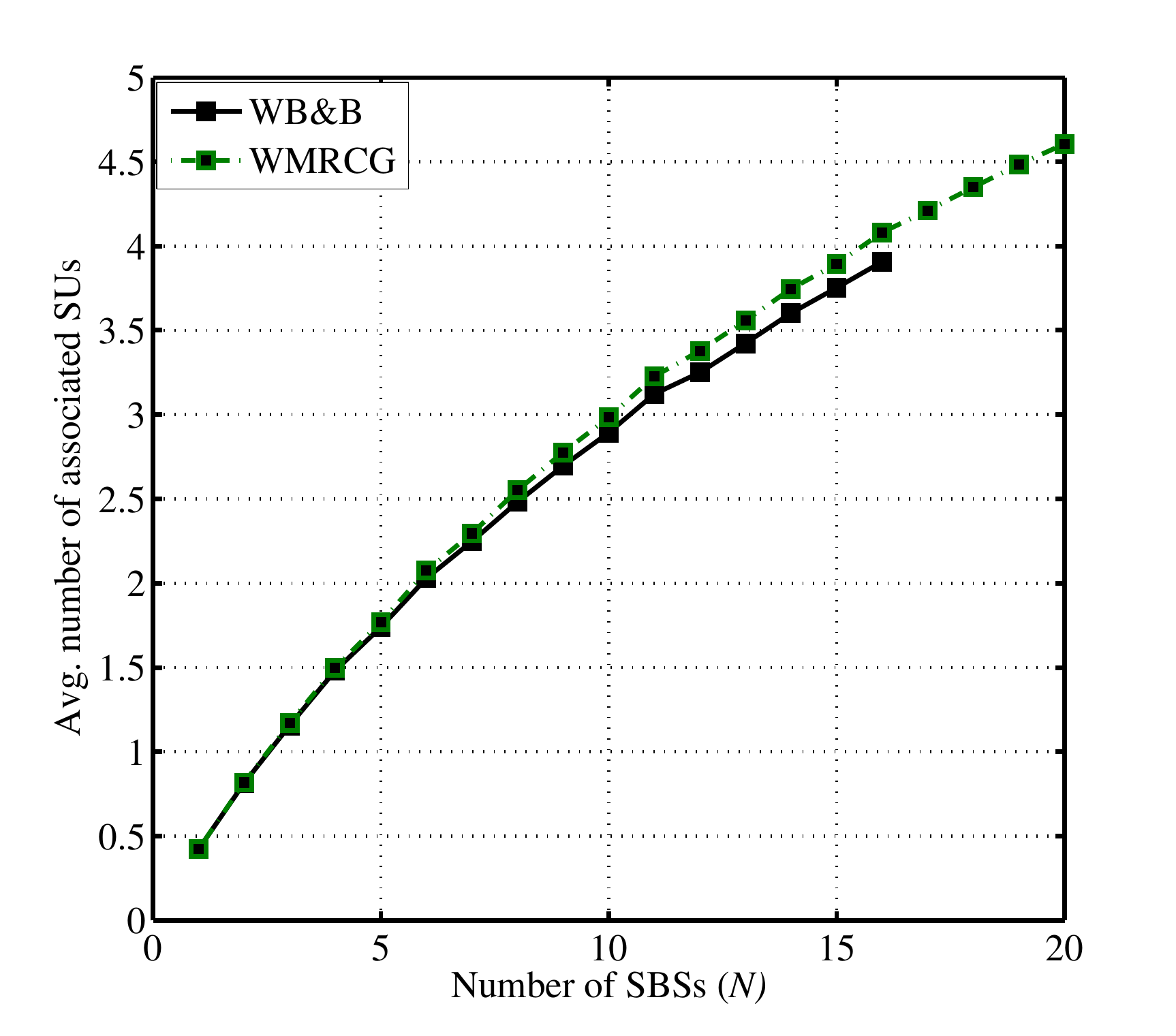}
		\caption{Performance of the WB\&B algorithm and the WMRCG algorithm for the weighted user-BS association.}
		\label{opt_grd_w}
	\end{figure}
	Fig.~\ref{opt_grd_w} considers the proposed WMRCG solution and B\&B solution for the weighted user-BS association problem \eqref{w:pbmatrixform}, denoted WB\&B. It can be seen that WMRCG algorithm provides slightly higher number of associated SUs than WB\&B algorithm since the latter does not maximize the number of associated SUs but it maximizes a weighted sum of associated SUs as can be seen by the objective function given in \eqref{eq:costmw}. Comparing Fig.~\ref{opt_grd_u} and Fig.~\ref{opt_grd_w}, it can be seen that the weighted solution has less performance than the unweighted one in terms of average number of associated SUs. This performance loss is compensated by gains in fairness as shown in the next simulations.
	
	\begin{figure}[h!]
		\centering
		\includegraphics[scale=0.8]{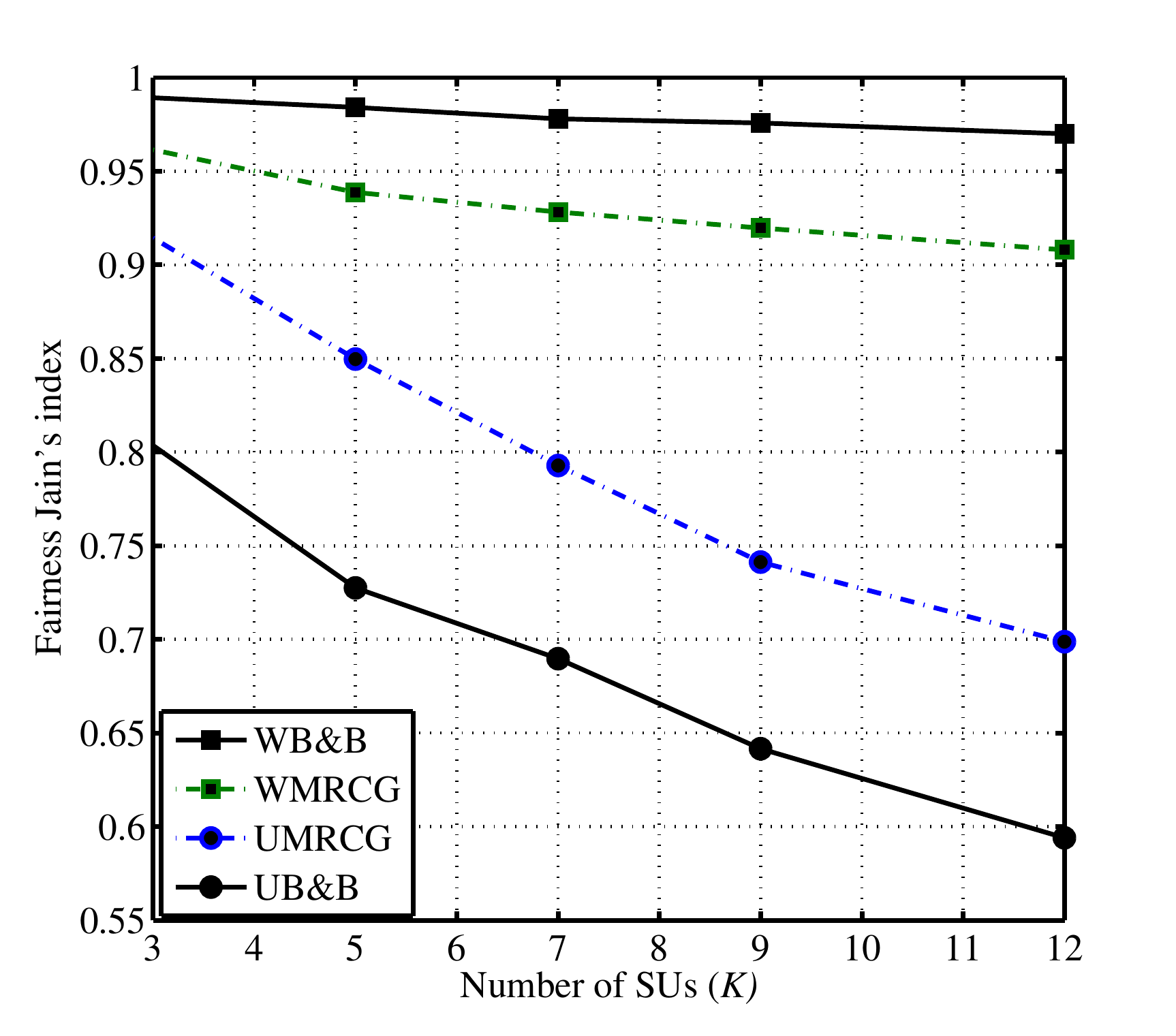}
		\caption{Comparison of the UB\&B algorithm, the WB\&B algorithm, the UMRCG algorithm and the WMRCG algorithm in terms of the fairness between SUs. $N = 6$.}
		\label{fair}
	\end{figure}
	To measure the fairness, the location of SBSs is assumed fixed whereas the SUs are located randomly with uniform distribution in the network. The fairness measure used in the simulations is the well-known Jain's index \cite{jain}.
	
	Fig. \ref{fair} demonstrates the fairness of the proposed algorithms along with the optimal ones. The weights are obtained by equation \eqref{we2}. We observe that WB\&B gives the highest fairness index. Also, WMRCG gives a high fairness index. On the other hand, UMRCG and UB\&B produce the worst results of fairness index as expected. We also see that when the number of SUs increases, the network starts to densify, and the fairness of all algorithms suffer.
	
	\begin{figure}[h!]
		\centering
		\includegraphics[scale=0.8]{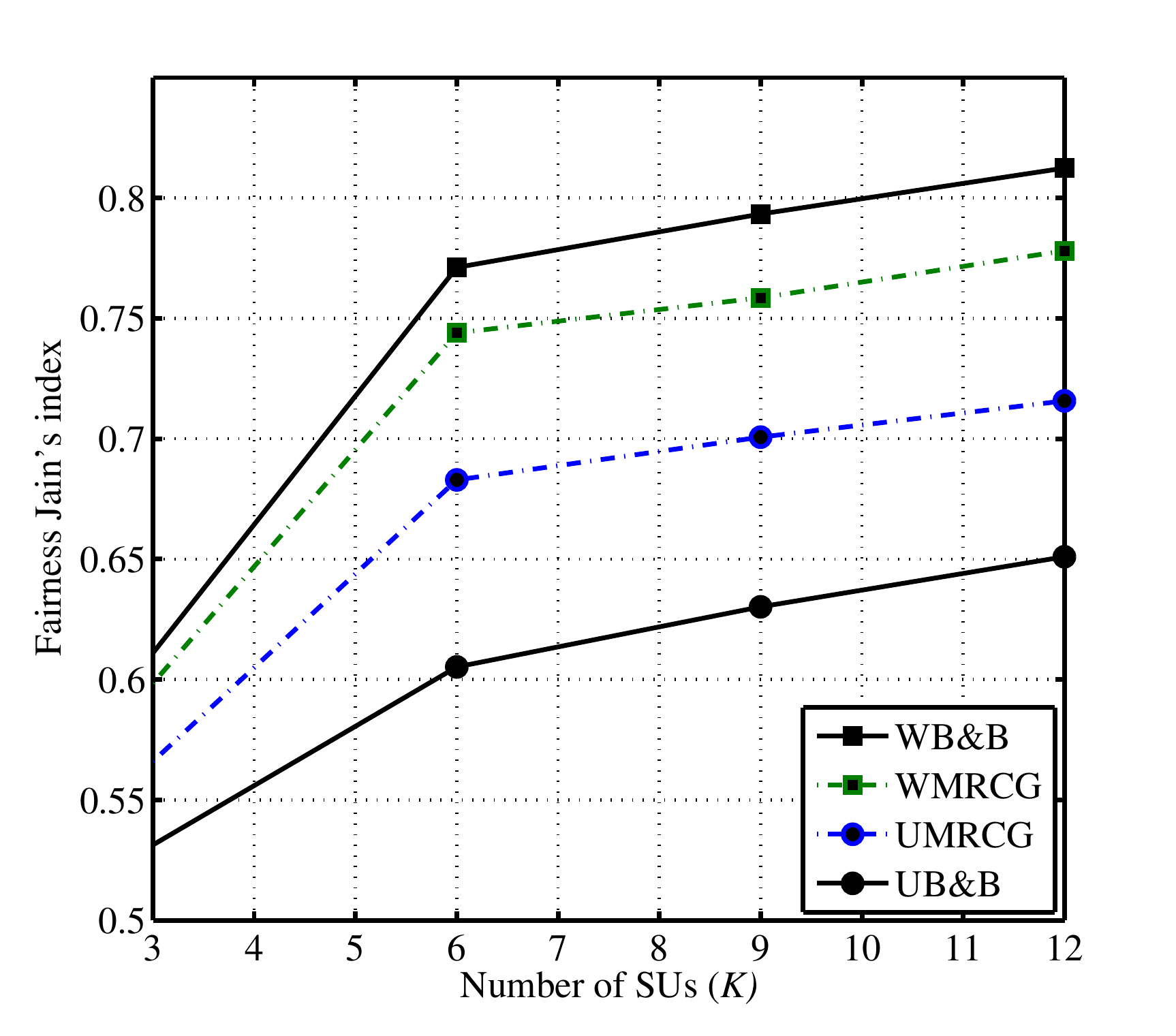}
		\caption{Comparison of the UB\&B algorithm, the WB\&B algorithm, the UMRCG algorithm and the UMRCG algorithm in terms of the load balancing of SBSs. $N = 6$.}
		\label{fair_f}
	\end{figure}
	Fig. \ref{fair_f} shows the fairness between SBSs of the proposed algorithms. As discussed is Section~\ref{prob}, the fairness between the SBSs is considered as a load balancing between the cells. The weights are obtained by equation \eqref{we3}. It is clear that as long as the number of SUs in the network is large, the load among different SBSs is balanced since more opportunities are given to each SBS to be associated. Further, WB\&B and WMRCG still give the best results in terms of fairness between SBSs compared to UB\&B and UMRCG.
	\clearpage
	\section{Conclusion\label{cl}}
	
	This paper studies the problem of user-BS association in a HetSNet of co-channel densely deployed SBSs and a MBS. The user-BS association problem is modeled as a linear integer program. The objective is to maximize the number of associated SUs subject to QoS constrains defined by SINR. This paper proves that the unweighted user-BS association problem is NP-hard. Then, two heuristic algorithms are proposed, namely the UMRCG algorithm and the WMRCG algorithm. Next the complexity of the proposed algorithms are derived and shown to be polynomial in time. The performance of the proposed algorithms are compared against the optimal exponential-time BF and B\&B algorithms. Moreover the performance is also compared against the max-SINR algorithm and a recently proposed algorithm in~\cite{MingyiH}. The proposed algorithms outperforms all previously proposed algorithms and is close to the optimal solution as demonstrated by simulations.
	
	The future extensions of this research will propose algorithms for joint power control and user-BS association and study the effect of statistical knowledge of channel information on the performance of these algorithms. Also, we will study the hardness of approximating the user-BS association problem and develop approximation algorithms with provable guarantees. Furthermore, fully distributed algorithms will be developed to solve the user-BS association problem using game theory and machine learning.
	
	\section*{Acknowledgement}
	
	The authors would like to thank Dr. Elmahdi Driouch for his valuable comments and suggestions.
	
	\clearpage
	\appendices
	
	\section{Proof of Lemma~\ref{lm}}
	\label{appendixA}
	
	\begin{IEEEproof}
		We prove that \textsc{Set Cover} $\propto$ \textsc{0-Valid Max Ones}. 
		Let $I_{sc}$ and $I_{0v}$ be two respective instances of \textsc{Set Cover} problem and the \textsc{0-Valid Max Ones} problem which are given by: $I_{sc}=\left(\mathcal{U}, \{\{S_j\}\,\forall\,j\}, k\right),$ and $I_{0v}=\left(f(x_1, \cdots, x_n)\right)$. 
		
		From the instance $I_{sc}$, we construct the instance $I_{0v}$ as follow. From each subset $S_j$ of $\mathcal{J}$, the matrix $\mathbf{M}=\bigl[m_{xy}\bigr]_{x\in\{1,\ldots,X\},y\in\{1,\ldots,Y\}}$, where $X=|\mathcal{J}|$ and $Y=\max\left\{\bigcup_j S_j\right\}$ is constructed as follows:
		
		\begin{align}
		\label{mm}
		m_{xy}\myeq
		\begin{cases}
		y & \text{if $y \in S_x$},\\ 
		0 & \text{otherwise.}
		\end{cases}
		\end{align}
		Based on the steps given by algorithm~\ref{sc20valid} and using the matrix defined by equation \eqref{mm}, the instance $I_{0v}$ is easily obtained.
		
		\begin{algorithm}
			\DontPrintSemicolon
			\KwIn{An instance of \textsc{Set Cover} $\left(\mathcal{U},\mathcal{J}=\{S_j\}, k\right)$.}
			\KwOut{An instance of \textsc{0-Valid Max Ones}.}
			Construct the matrix $\mathbf{M}$ according to~\eqref{mm}\;
			\For{$j=1$ \textup{to} $Y$}
			{
				$C_j=1$\;
				\For{$i=1$ \textup{to} $X$}
				{
					\If{$m_{ij}\neq 0$}
					{
						$C_j=C_j\vee\neg\,x_{i}$\;
					}
				}
			}
			$f(x_1,\cdots,x_{X})=C_1\land C_2\land\ldots\land C_{Y}$\;
			\Return{An instance of \textsc{0-Valid Max Ones}.}\;
			\caption{{\sc \textsc{Set Cover to 0-Valid Max Ones}}}
			\label{sc20valid}
		\end{algorithm}
		Finally, if \textsc{0-Valid Max Ones} is solved with the instance $I_{0v}$ then the optimal solution $\mathbf{x}=(x_1, \cdots, x_n)$ contains the least possible number of zeros. Let $I$ be the set of zeros in the solution $\mathbf{x}$. Thus, the solution of the \textsc{Set Cover} problem corresponds to the subfamily of sets $\mathcal{I}=\{\{S_{p}\},\,\forall\,p\in I\}$. Hence, \textsc{Set Cover} problem is solved with the minimum number of subsets. The reduction from \textsc{Set Cover} to \textsc{0-Valid Max Ones} is done in polynomial time as illustrated in Algorithm~\ref{sc20valid}. Therefore, \textsc{Set Cover} $\propto$ \textsc{0-Valid Max Ones} which proves Lemma~\ref{lm}.
	\end{IEEEproof}
	
	\section{Proof of Theorem~\ref{thm}}
	\label{appendixB}
	
	\begin{IEEEproof}
		We show that \textsc{0-Valid Max Ones} $\propto$ problem \eqref{unw:pbmatrixform}. Let $I_1=(K, N, \mathbf{A})$ be an instance of problem \eqref{unw:pbmatrixform} where $K$ is the number of SUs, $N$ is the number of SBSs, $\mathbf{A}$ is the matrix defined in problem \eqref{unw:pbmatrixform}. Let $I_2=(f(x_1, \cdots, x_n))$ be an instance of the \textsc{0-Valid Max Ones} problem.
		
		An instance of problem \eqref{unw:pbmatrixform} can be constructed by converting the set of Boolean clauses of the binary constraint $f(\cdot)$ to a system of linear inequalities. Therefore, $f(\cdot)$ is true $\Leftrightarrow\mathbf{A} \mathbf{x} \leqslant \mathbf{1}$. Hence, the problem of maximizing the number of associated SUs while the SINR requirements are met (i.e., $\mathbf{A} \mathbf{x} \leqslant \mathbf{1}$) is equivalent to the problem of maximizing the number of true literals while the Boolean formula is true (i.e., $f(\cdot)$ is true).
		
		In order to get the instance $I_1$ from the instance $I_2$, the following transformation is applied. First, let $\mathcal{S}_l=\{i_1^l, \dotsc, i_k^l\}$ be a subset of $\{1, \dotsc, n\}$ for some $l\in L$ and some $k\in\{1, 2, \dotsc, n\}$. Then, for each clause $l$ of $f(\cdot)$, i.e., $\,\bigvee_{i \in \mathcal{S}_l}\neg\,x_i$, the following system of linear inequalities is given: 
		\begin{align*}
		\gamma\,g_{i_{\sigma}^l\,i_{\sigma}^l}<\left(\sum\limits_{\substack{p=1 \\ p\neq\,\sigma}}^{k}\gamma\,g_{i_{\sigma}^l\,i_{p}^l}+1\right)\beta,\;\forall\,\sigma\in\{1, \dotsc, k\}.
		\end{align*}
		Second, this system of linear inequalities is easily solved in polynomial time to get the corresponding $g_{ij}$ since it has more many variables than equations. Let $\mathcal{A}$ and $\mathcal{B}$ be the sets of solutions of the $g_{ij}$. Then, the remainder values of $g_{ij}$ will be set to $0$, i.e., $g_{ij}=0,\,\forall\,i\not\in \mathcal{A}, \forall\,j\not\in \mathcal{B}$. Using this transformation, we can get the matrix $\mathbf{A}$, $K$, and $N$ where $K=N=Y$. Therefore an instance of problem \eqref{unw:pbmatrixform} is obtained. Finally, if problem \eqref{unw:pbmatrixform} is solved using $I_1$ and let the solution vector be $\mathbf{x}$, then, if $x_{ij}=1\Leftrightarrow\,i=j$ and the corresponding Boolean variable is set to $1$. Hence, in the instance $I_2$, we have $x_i=x_j=1$ and $x_{i'}=0,\,\forall\,i'\neq i,j$. Therefore, \textsc{0-Valid Max Ones} is solved. At last, we can verify easily in polynomial time that the constraints evaluate true is equivalent to the Boolean formula evaluates true. Therefore, problem \eqref{unw:pbmatrixform} is solved if and only if \textsc{0-Valid Max Ones} is solved. 
		
		To conclude, \textsc{0-Valid Max Ones} $\propto$ problem \eqref{unw:pbmatrixform} and therefore the unweighted user-BS association problem is NP-hard. This proves Theorem~\ref{thm}.
	\end{IEEEproof}
	
	\bibliographystyle{IEEEtran}
	\bibliography{IEEEabrv,bibliotex}

\begin{thebibliography}{10}
\providecommand{\url}[1]{#1}
\csname url@samestyle\endcsname
\providecommand{\newblock}{\relax}
\providecommand{\bibinfo}[2]{#2}
\providecommand{\BIBentrySTDinterwordspacing}{\spaceskip=0pt\relax}
\providecommand{\BIBentryALTinterwordstretchfactor}{4}
\providecommand{\BIBentryALTinterwordspacing}{\spaceskip=\fontdimen2\font plus
\BIBentryALTinterwordstretchfactor\fontdimen3\font minus
  \fontdimen4\font\relax}
\providecommand{\BIBforeignlanguage}[2]{{%
\expandafter\ifx\csname l@#1\endcsname\relax
\typeout{** WARNING: IEEEtran.bst: No hyphenation pattern has been}%
\typeout{** loaded for the language `#1'. Using the pattern for}%
\typeout{** the default language instead.}%
\else
\language=\csname l@#1\endcsname
\fi
#2}}
\providecommand{\BIBdecl}{\relax}
\BIBdecl

\bibitem{interference_management}
T.~Zahir, K.~Arshad, A.~Nakata, and K.~Moessner, ``Interference management in
  femtocells,'' \emph{{IEEE} Commun. Surveys Tuts.}, vol.~15, no.~1, pp.
  293--311, Q1 2013.

\bibitem{ekram}
E.~Hossain, L.~B. Le, and D.~Niyato, \emph{Self-Organizing Small Cell
  Networks}.\hskip 1em plus 0.5em minus 0.4em\relax John Wiley \& Sons, Inc.,
  2013.

\bibitem{andrews_survey}
V.~Chandrasekhar, J.~Andrews, and A.~Gatherer, ``Femtocell networks: a
  survey,'' \emph{{IEEE} Commun. Mag.}, vol.~46, no.~9, pp. 59--67, Sept. 2008.

\bibitem{5g}
J.~Andrews, S.~Buzzi, W.~Choi, S.~Hanly, A.~Lozano, A.~Soong, and J.~Zhang,
  ``What will 5g be?'' \emph{{IEEE} J. Sel. Areas Commun.}, vol.~32, no.~6, pp.
  1065--1082, June 2014.

\bibitem{femto_past}
J.~Andrews, H.~Claussen, M.~Dohler, S.~Rangan, and M.~Reed, ``Femtocells: Past,
  present, and future,'' \emph{{IEEE} J. Sel. Areas Commun.}, vol.~30, no.~3,
  pp. 497--508, Apr. 2012.

\bibitem{Kuang}
Q.~Kuang, J.~Speidel, and H.~Droste, ``Joint base-station association, channel
  assignment, beamforming and power control in heterogeneous networks,'' in
  \emph{Proc. IEEE Veh. Technol. Conf. (VTC Spring)}, Yokohama, Japan, 6-9 May
  2012.

\bibitem{Qian}
L.~P. Qian, Y.~J. Zhang, Y.~Wu, and J.~Chen, ``Joint base station association
  and power control via benders' decomposition,'' \emph{{IEEE} Trans. Wireless
  Commun.}, vol.~12, no.~4, pp. 1651--1665, Apr. 2013.

\bibitem{andrews_load}
Q.~Ye, B.~Rong, Y.~Chen, M.~Al-Shalash, C.~Caramanis, and J.~Andrews, ``User
  association for load balancing in heterogeneous cellular networks,''
  \emph{{IEEE} Trans. Wireless Commun.}, vol.~12, no.~6, pp. 2706--2716, Jun.
  2013.

\bibitem{load_assoc}
K.~Son, S.~Chong, and G.~de~Veciana, ``Dynamic association for load balancing
  and interference avoidance in multi-cell networks,'' in \emph{Proc. IEEE Int.
  Symp. on Modeling and Optimization in Mobile, Ad Hoc and Wireless Networks
  and Workshops (WIOPT)}, Limassol, Cyprus, 16-20 Apr. 2007.

\bibitem{KaimingS}
K.~Shen and W.~Yu, ``Distributed pricing-based user association for downlink
  heterogeneous cellular networks,'' \emph{{IEEE} J. Sel. Areas Commun.},
  vol.~32, no.~6, pp. 1100--1113, June 2014.

\bibitem{Gibbs}
C.~S. Chen and F.~Baccelli, ``Self-optimization in mobile cellular networks:
  Power control and user association,'' in \emph{Proc. IEEE Int. Conf. on
  Communications (ICC)}, Cape Town, South Africa, 23-27 May 2010.

\bibitem{Olga}
O.~Goussevskaia, Y.~A. Oswald, and R.~Wattenhofer, ``Complexity in geometric
  {SINR},'' in \emph{Proc. ACM Int. Symp. on Mobile Ad Hoc Networking and
  Computing (MOBIHOC)}, Montreal, QC, Canada, 9-14 Sept. 2007.

\bibitem{5062048}
M.~Andrews and M.~Dinitz, ``Maximizing capacity in arbitrary wireless networks
  in the {SINR} model: Complexity and game theory,'' in \emph{Proc. IEEE Int.
  Conf. on Computer Communications (INFOCOM)}, Rio de Janeiro, Brazil, 19-25
  Apr. 2009.

\bibitem{Dinitz:2010:DAA:1833515.1833717}
M.~Dinitz, ``Distributed algorithms for approximating wireless network
  capacity,'' in \emph{Proc. IEEE Int. Conf. on Computer Communications
  (INFOCOM)}, San Diego, CA, USA, 15-19 Mar. 2010.

\bibitem{MingyiH}
R.~Sun, M.~Hong, and Z.-Q. Luo, ``Joint downlink base station association and
  power control for max-min fairness: Computation and complexity,''
  \emph{{IEEE} J. Sel. Areas Commun.}, vol.~33, no.~6, pp. 1040--1054, June
  2015.

\bibitem{Crescenzi200110}
P.~Crescenzi, R.~Silvestri, and L.~Trevisan, ``On weighted vs unweighted
  versions of combinatorial optimization problems,'' \emph{Information and
  Computation}, vol. 167, no.~1, pp. 10 -- 26, 2001.

\bibitem{Rappaport:2001:WCP:559977}
T.~Rappaport, \emph{Wireless Communications: Principles and Practice},
  2nd~ed.\hskip 1em plus 0.5em minus 0.4em\relax Upper Saddle River, NJ, USA:
  Prentice Hall PTR, 2001.

\bibitem{zvi}
C.~Avin, Z.~Lotker, and Y.-A. Pignolet, ``On the power of uniform power:
  Capacity of wireless networks with bounded resources,'' in
  \emph{Algorithms-ESA 2009}.\hskip 1em plus 0.5em minus 0.4em\relax Springer
  Berlin Heidelberg, 2009, vol. 5757, pp. 373--384.

\bibitem{Schrijver:1986:TLI:17634}
A.~Schrijver, \emph{Theory of Linear and Integer Programming}.\hskip 1em plus
  0.5em minus 0.4em\relax New York, NY, USA: John Wiley \& Sons, Inc., 1986.

\bibitem{5567009}
M.~Mehrjoo, M.~Awad, M.~Dianati, and X.~Shen, ``Design of fair weights for
  heterogeneous traffic scheduling in multichannel wireless networks,''
  \emph{{IEEE} Trans. Commun.}, vol.~58, no.~10, pp. 2892--2902, Oct. 2010.

\bibitem{doi:10.1137/S0097539799349948}
S.~Khanna, M.~Sudan, L.~Trevisan, and D.~Williamson, ``The approximability of
  constraint satisfaction problems,'' \emph{SIAM Journal on Computing},
  vol.~30, no.~6, pp. 1863--1920, Jun. 2001.

\bibitem{karp}
R.~Karp, ``Reducibility among combinatorial problems,'' in \emph{Complexity of
  Computer Computations}, R.~Miller, J.~Thatcher, and J.~Bohlinger, Eds.\hskip
  1em plus 0.5em minus 0.4em\relax Springer US, 1972, pp. 85--103.

\bibitem{comb}
E.~A. Bender, ``Asymptotic methods in enumeration,'' \emph{SIAM Review},
  vol.~16, no.~4, pp. 485--515, Apr. 1974.

\bibitem{cplex}
``{IBM ILOG CPLEX Optimizer},''
  \url{http://www-01.ibm.com/software/integration/optimization/cplex-optimizer/},
  Dec. 2010, accessed: 2013-01-24.

\bibitem{olga:2}
O.~Goussevskaia, M.~M. Halld\'{o}rsson, and R.~Wattenhofer, ``Algorithms for
  wireless capacity,'' \emph{IEEE/ACM Trans. Netw.}, vol.~22, no.~3, pp.
  745--755, 2014.

\bibitem{CW12a}
J.~Currie and D.~I. Wilson, ``{OPTI: Lowering the Barrier Between Open Source
  Optimizers and the Industrial MATLAB User},'' in \emph{{Foundations of
  Computer-Aided Process Operations}}, N.~Sahinidis and J.~Pinto, Eds.,
  Savannah, Georgia, USA, Jan. 2012.

\bibitem{jain}
R.~K. Jain, D.-M.~W. Chiu, and W.~R. Hawe, ``{A Quantitative Measure Of
  Fairness And Discrimination For Resource Allocation In Shared Computer
  Systems},'' DEC-TR-301, Digital Equipment Corporation, Tech. Rep., Sept.
  1984.

\end{thebibliography}
	
\end{document}